\pdfoutput=1   
\documentclass[usenatbib,usegraphicx,fleqn]{ctextemp_mn2e}
\usepackage{amsmath,amssymb,bm,color,multirow,multicol}
\usepackage{charter}   

\newcommand{\e}[1]{\times 10^{#1}}

\usepackage{ulem}

\title[MRI-driven Accretion on to Magnetized stars:]{MRI-driven Accretion on to Magnetized stars:
Global 3D MHD Simulations of Magnetospheric and Boundary Layer Regimes}

\author[M. M. Romanova et al.]
{M. M. Romanova,$^1$\thanks{e-mail:romanova@astro.cornell.edu}, G.
V. Ustyugova,$^2$,
A. V. Koldoba$^2$, R. V. E. Lovelace $^{1}$\\
$^1$ Department of Astronomy, Cornell University, Ithaca, NY 14853-6801, USA\\
$^2$ Keldysh Institute for Applied Mathematics, Moscow, Russia\\}

\begin{document}

\maketitle

\begin{abstract}

\noindent We discuss results of global three-dimensional (3D)
magnetohydrodynamic (MHD) simulations of accretion on to a rotating
magnetized star with a tilted dipole magnetic field, where the
accretion is driven by the \textit{magneto-rotational instability} (MRI).
   The simulations show that MRI-driven turbulence develops in the
disc, and angular momentum  is transported outwards due primarily
to the magnetic stress.  The turbulent flow is strongly
inhomogeneous and the densest matter is in azimuthally-stretched
turbulent cells. We investigate two regimes
of accretion: a magnetospheric regime and  a boundary layer (BL) regime.
 In the magnetospheric regime, the magnetic field of the star is dynamically important:  the accretion disc is truncated by the star's magnetic
 field within a few stellar radii from the star's surface, and matter
flows to the star in funnel streams.
   The funnel streams flow towards the
south and north magnetic poles but are not equal due to the inhomogeneity
of the flow.     The hot spots on the stellar surface are not symmetric as well.
     In the BL regime, the magnetic field of the star is
dynamically unimportant, and matter accretes to the surface of the
star through the boundary layer.
       The magnetic
field in the inner disc is strongly amplified by the shear of the
accretion flow, and the matter and magnetic stresses become
comparable.
   Accreting matter forms a belt-shaped hot region
on the surface of the star. The belt has inhomogeneous density distribution which varies in time
due to variable accretion rate.
   The peaks in the variability curve are associated with accretion of individual
turbulent cells. They show 20-50\% density amplifications at
periods $\sim 5-10$ dynamical time-scales
at the surface of the star.
Spiral waves in the disc are excited in both, magnetospheric and
boundary layer regimes of accretion. Results of simulations can be
applied to classical T Tauri stars, accreting brown dwarfs, millisecond pulsars, dwarf novae cataclysmic
variables, and other stars with magnetospheres smaller than several stellar
radii.

\end{abstract}

\section{Introduction}
\label{sec:introduction}

Different types of stars  have \textit{dynamically
important} magnetic fields. These include young, classical T Tauri
stars (hereafter - CTTS; e.g., \citealt{bouv07}), young accreting brown dwarfs (e.g., \citealt{moha05}),
accreting neutron
stars (e.g., \citealt{lewi95,vand00}),
and different types of cataclysmic
variables, including dwarf novae and intermediate polars  (e.g., \citealt{hell01,warn03}).
In such stars, the
accretion disc is truncated at the disc-magnetosphere boundary,
and the magnetic field governs the matter flow (e.g.,
\citealt{prin72,ghos79,koni91,coll93,wang95,camp10}). We refer to
this regime as the \textit{magnetospheric regime} of accretion. On
the other hand, many accreting neutron stars, white dwarfs, and other stars are
expected to have relatively weak, \textit{dynamically-unimportant}
magnetic field, and accrete in the equatorial region of the star
in the \textit{boundary layer regime} (e.g.,
\citealt{poph95,inog99,piro04,fisk05}).

The physics of the disc-magnetosphere interaction depends on both the
properties of the magnetized star and the properties of the accretion disc. Accretion
in the disc can be smooth as modelled by including an $\alpha-$viscosity
\citep{shak73} in the MHD equations. Or, it can be strongly turbulent owing to the
growth of the  magneto-rotational instability \citep{veli59,chan60,balb91,balb98}.

Interaction of a magnetized star with $\alpha-$type discs
has been investigated in a number of axisymmetric simulations
(e.g., \citealt{mill97,roma02,long05,bess08}), and
in global 3D simulations (e.g.,
\citealt{roma03,roma04,roma08,kulk05,kulk08,long07,long08,long11}).
These simulations confirmed many properties of the
disc-magnetosphere interaction predicted by theory.  They also
revealed new features, connected, for example, with the
non-stationarity of the disc-magnetosphere interaction and the
possibility of outflows (e.g.,
\citealt{love95,good97,good99,roma05,usty06,roma09}),
and possible important role of the interchange
instability  \citep{roma08,kulk08,kulk09}. Interaction of
a magnetized star with a turbulent accretion disc has not been
studied as much.
MRI-driven turbulence has been extensively studied in
axisymmetric and local/global 3D MHD simulations
(e.g., \citealt{hawl95,bran95,ston96,
armi98,hawl01,ston01,beck09,beck11,floc11,simo11}). However, in
all these simulations, the central object is a \textit{black
hole}.

Recently, we performed the first axisymmetric (2.5D) MHD
simulations of the MRI-driven accretion on to a \textit{magnetized star}
\citep{roma11a}.
We observed different phenomena connected with turbulent
nature of the disc. However, global 3D simulations are required
for more realistic description of MRI-driven turbulence and
magnetized stars.

 In this paper, we present results from  global
 3D MHD simulations of MRI-driven, turbulent
accretion on to a \textit{magnetized star} with a \textit{tilted} dipole
magnetic field.
      It is necessary to perform the simulations in the
full three-dimensional space, including the full range of the
azimuthal angle ($0-2\pi$). Furthermore, we need to have
sufficiently high grid resolution in the disc for the magnetic
turbulence to develop and be sustained, and we need to resolve the
magnetosphere of the star, which has strong gradients of the
dipole magnetic field. Because of these requirements, significant
computing time is required, and we performed simulations only
during a restricted period of time.

First, we show results of MRI-driven accretion on
to stars with dynamically unimportant magnetic
field, accreting in the BL regime. In this case, we were able to
obtain relatively long runs
 and investigated MRI-driven accretion in the disc.
We also studied the properties of the inner regions of the disc,
the disc-star boundary, and
variability associated with the turbulence.
Next, we show results for the magnetospheric accretion where
the disc matter is truncated by the magnetosphere. In this case, we concentrated on the disc-magnetosphere interaction and accretion through the funnel flow.

In \S 2 we describe the problem setup.
In \S 3 and \S 4  we show results of
simulations for accretion in the boundary layer and magnetospheric regimes, respectively.
In \S 5 we discuss the different stresses in the disc. \S 6 gives the conclusions from this work.

\section{Problem setup}
\label{sec:problem-setup}

\begin{figure}
\centering
\includegraphics[width=8cm]{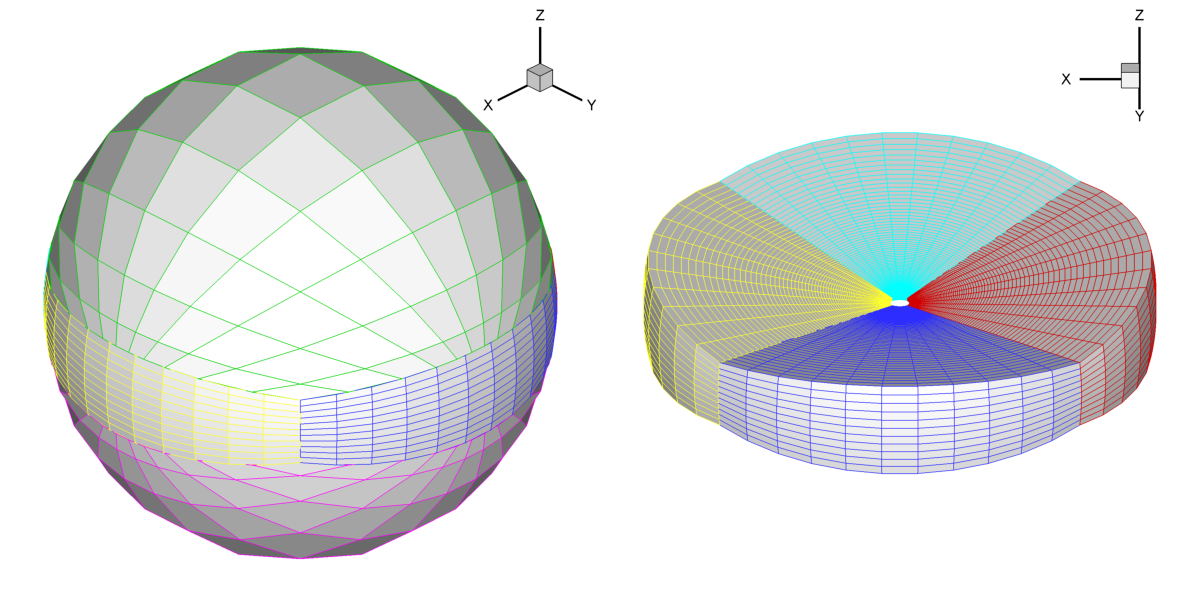}
\caption{The ``cubed sphere'' grid used in our simulations. The
grid is compressed towards the equatorial plane. A low resolution
case  is shown here to make the grid visible.} \label{cubed-comp}
\end{figure}

\begin{figure*}
\centering
\includegraphics[width=16cm]{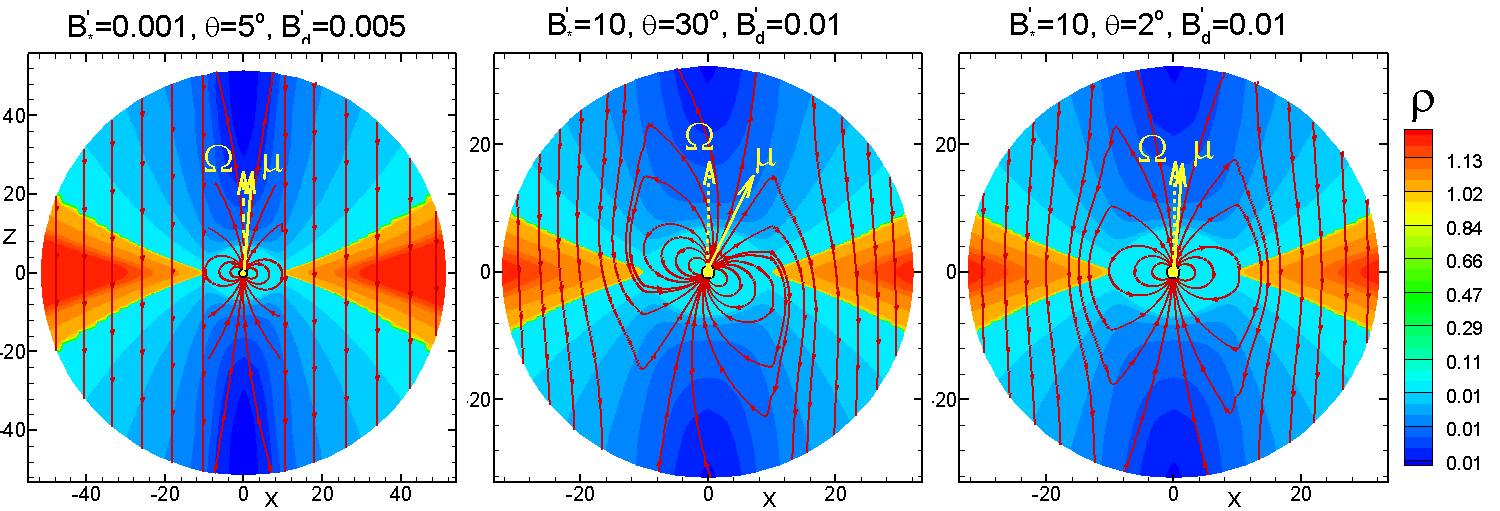}
\caption{The initial distribution of density (color background)
and magnetic field lines in the $xz-$slice. The left-hand
panel shows the case of a dynamically unimportant and slightly
tilted stellar magnetic field ($B'_\star=0.001$,
$\Theta=5^\circ$).
 The two right-hand panels show the cases of
 a dynamically important stellar
magnetic field  ($B'_\star=10$) at high ($\Theta=30^\circ$) and
low ($\Theta=2^\circ$) tilts of the dipole field.  The magnetic
field in the disc is $B'_d=0.005$ and $B'_d=0.01$, respectively.}
\label{init-3}
\end{figure*}

\begin{figure*}
\centering
\includegraphics[width=14cm]{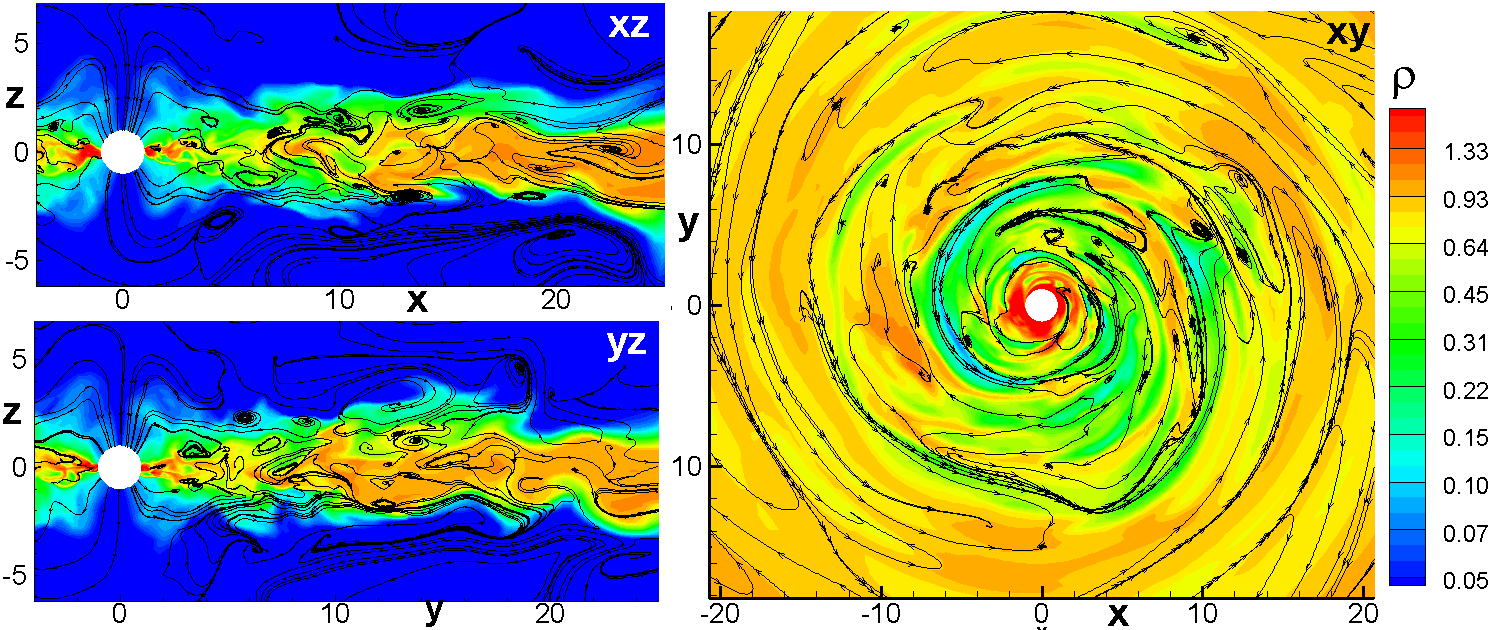}
\caption{Slices of the density distribution (color background) and
sample field lines in the case of a star with dynamically
unimportant magnetic field ($B'_d=0.001$, $\Theta=5^\circ$). The
left-hand panels show the $xz$ and $yz$ slices, while the
right-hand panel shows an equatorial,  $xy$, slice.}
\label{d001-t5-slice-3-vac}
\end{figure*}

\subsection{Magnetosphere and magnetospheric radius}
\label{sec:setup-magnetospheric}

If the magnetic field of the star is relatively strong
 (in the sense of being dynamically important)\footnote{Here, and below in the paper, we
use terms `strong' and `weak' for dynamically important and dynamically unimportant fields.},
then the accretion disc is truncated by the stellar magnetosphere at some
radius $r_m$, which is called the \textit{magnetospheric radius}. Different physical
arguments were proposed to define this radius (e.g., \citealt{prin72,ghos79}). For example, this radius
can be derived from the balance between the magnetic pressure of the magnetosphere
and the ram pressure of the accreting
matter (e.g., \citealt{lamb73}): $B^2/8\pi\approx\rho v^2$
(where $B$ is the local value of the magnetic field, $\rho$ and $v$ are density and total velocity in the disc).
 Then, the magnetospheric radius can be derived in the form, which is often used in theoretical analysis
 (e.g, \citealt{lamb73,koni91}):
\begin{equation}
r_m=k \mu^{4/7} {\dot M}^{-2/7}{GM}^{-1/7},
\label{eq:r_m}
\end{equation}
where $\mu=B_\star R_\star^3$ is the magnetic moment of the star, $\dot M$ is the accretion rate from the disc,
and coefficient $k\approx 0.5$ \citep{long05}.

In simulations, however, we do not know the accretion rate in advance,  and it varies with time.
Hence, it is better to find the magnetospheric radius
 directly from the balance of stresses.  We introduce
\begin{equation}
\beta_1= \frac{p + \rho
v_\phi^2}{B^2/8\pi}~,~~~~~~\beta=\frac{p}{B^2/8\pi}~,
\label{eq:beta1-beta}
\end{equation}
where, $p$ is the gas pressure, $v_\phi$ is the azimuthal component of the magnetic field.    Here, $\beta$
is the commonly used plasma parameter.
The modified plasma parameter
$\beta_1$ takes into account the matter stress
 associated the plasma flow in the disc
and it is often more relevant in astrophysical situations than $\beta$ parameter \citep{roma02}.
In accretion discs, the magnitude flow velocity is approximately
the azimuthal velocity, $v\approx v_\phi$,
and hence the matter stress has also a meaning of the ram pressure, and the
condition $\beta_1=1$ can be also interpreted as a balance of total pressure at the disc-star boundary.
      We observed from simulations that this condition
always gives the boundary between the disc and the magnetosphere.
 We use this condition to find $r_m$ from simulations.
Typically we have $\beta\approx 1$ at radii larger than $r_m$,
which often shows the place where the funnel streams start flowing from
the disc to the star (see also \citealt{bess08,camp10}).
\footnote{\citet{bess08} compared several approaches for derivation of $r_m$ and showed  that this radius
does not differ much. This is
because the magnetic pressure $\sim B^2\sim r^{-6}$ is a steep
function of $r$.}.

Modeling of accreting magnetized stars is technically challenging,
 because in regions of the strongest magnetic field, the
matter density is low, and this requires a small time-step
in simulations.
     In $\alpha-$type discs, we were able to model
relatively large magnetospheres with $r_m \lesssim 10 R_\star$ (e.g., \citealt{roma06}).
     However, in the
current simulations of turbulent discs, we model somewhat smaller
magnetospheres with $r_m\approx 3 R_\star$, because the duration of simulations strongly increases with the size of the magnetosphere.
There are several types of stars where such magnetospheres are expected (e.g.,
CTTSs, accreting millisecond pulsars, dwarf novae, accreting brown
dwarfs, and other stars).

In case of much weaker, dynamically unimportant field, the star's magnetosphere does not truncate the disc.
   That is, the  magnetospheric radius is equal or smaller than the radius of the star,  $r_m\lesssim R_\star$.
 We use this case for investigation of
the MRI-driven accretion in the disc and for analysis of processes at the disc-star boundary in the presence of the magnetic field.
    These simulations are faster, and they also
correspond to an important astrophysical regime of accretion
through the boundary layer.

\begin{figure*}
\centering
\includegraphics[width=12.0cm]{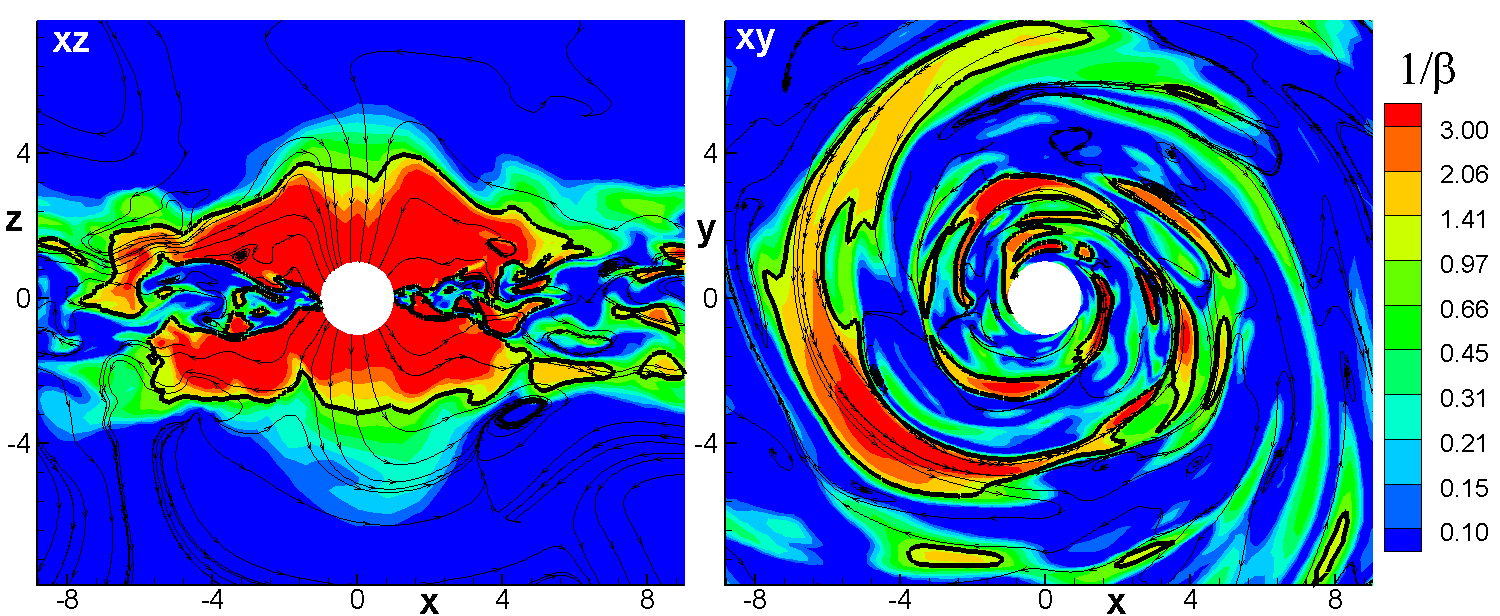}
\caption{The color background shows $x-z$ (left panel) and $x-y$ (right panel) slices of the magnetic pressure normalized to matter pressure, $\beta^{-1} =B^2/8\pi p$.
Thin lines are magnetic field lines, and the bold lines correspond to $\beta=1$.} \label{d001-t5-beta-2}
\end{figure*}

\subsection{Rotation of the star and corotation radius}
\label{sec:setup-rotation}

Another important parameter of the problem is the corotation radius,
$r_{\rm cor}=(GM_\star/\Omega_\star^2)^{1/3}$, where
Keplerian angular velocity in the disc equals  the angular velocity of the star, $\Omega_\star$.
 The result of the
disc-magnetosphere interaction depends on the ratio between $r_m$
and $r_{\rm cor}$. In case of slow rotation, $r_{\rm cor}\gtrsim r_m$, the
magnetosphere spins the inner parts of the disc down, the
gravitational force is larger than the centrifugal force, and accretion is
favorable. In the opposite case of rapidly rotating star, $r_{\rm cor}
< r_m$, the magnetosphere transfers its angular momentum to the
matter of the inner disc, and a star is in the `propeller' regime,
where accretion is suppressed (e.g.,
\citealt{illa75,love99,roma05,usty06}). In this paper we
consider only the case of slowly rotating stars, which is
favorable for accretion. We chose rather large corotation radius,
$r_{\rm cor}=10 R_\star$, to be sure that we are always in the
accretion regime from the beginning of the simulations
\footnote{Initially, the inner disc radius is located at
$r=10R_\star$, and this fact determines our choice for
$r_{\rm cor}$. Smaller values of $r_{\rm cor}$ would be also sufficient.}. The ratio $r_{\rm cor}/r_m\approx
3$ is larger than that expected in the rotational
equilibrium state\footnote{In the rotational equilibrium state,
the torques on the star associated with spinning-up and
spinning-down are balanced.} where $r_{\rm cor}\approx (1.2-1.6)r_m$
(the coefficient in front of $r_m$ depends on parameters of the
model \citep{long05}). However, our earlier simulations of $\alpha-$discs, performed at
different ratios of $r_m/r_{\rm cor}$, show that the result is very
similar in cases where $r_{\rm cor}$ is \textit{larger}, or \textit{much larger} than
$r_m$. In both cases, the gravitational force dominates  and the
accretion through the funnel flow has been observed (e.g.,
\citealt{roma02,long05}).

\subsection{MRI-driven turbulence}
\label{sec:setup-mri-theory}

The turbulence in the disc is initiated and supported by the
magneto-rotational instability. Here, we briefly summarize the condition for the onset of
this instability (Balbus \& Hawley 1991) for a \textit{simple
case} where an axial magnetic field $B_0\hat{\bf z}$ threads a thin
Keplerian disc which rotates with  angular velocity
 $\Omega=(GM_\star/r^3)^{1/2}$.
      For axisymmetric perturbations of the disc with $\delta{\bf v}=
[\delta v_r(z,t), \delta v_\phi(z,t),0]$ and $\delta {\bf
B}=[\delta B_r(z,t),\delta B_\phi(z,t),0] $ and for perturbations
proportional to $\exp(ik_z z-i\omega t)$, one finds the dispersion
relation
\begin{equation}
\omega_\pm^2 = (k_z v_A)^2 +{1\over 2}\kappa_r^2 \pm \left[{1\over
4}\kappa_r^4+4(k_z v_A\Omega)^2\right]^{1/2}~,
\end{equation}
 where $v_A \equiv B_0/\sqrt{4\pi \rho}$
 is the Alfv\'en velocity and $\kappa_r \equiv [4\Omega^2 +2r\Omega
d\Omega/dr]^{1/2}$ is the radial epicyclic frequency of the disc.
In order for the perturbation to fit within the vertical extent of
the disc  one needs $k_z h \gtrsim 1$, where $h = c_s/\Omega$ is
the half-thickness of the disc and $c_s$ is the midplane isothermal sound
speed in the disc.

The instability will develop if $\omega_-^2 <0$ which happens if
$(kv_A)^2 < -2r\Omega d\Omega/dr$. For a Keplerian disc this
corresponds to $(kv_A)^2 <3\Omega^2$.
    Therefore, the above-mentioned condition that $k_z h
\gtrsim 1$ implies that the instability occurs only for $ v_A <
c_s$,  or the condition for plasma parameter becomes:
\begin{equation}
\beta = \frac{p}{B_0^2/{8\pi}}= {2 c_s^2 \over v_A^2} >1~,
\label{eq:beta-theor}
\end{equation}
where matter pressure $p=\rho c_s^2$ . Note that $\beta$ is based
on the {\it initial} vertical magnetic field, $B_0$. As a result
of the instability the magnetic field may grow to values much
larger than $B_0$.
    The maximum value of the growth
rate  is $\Im(\omega)_{\rm max}=3\Omega/4$, and it occurs for
$k_{\rm max}=(15/16)^{1/2}\Omega/v_A$.    For $\beta <1$ the
perturbation does not fit inside the disc and there is stability.
   As $\beta$ increases from unity (weaker magnetic field) the
maximum growth rate stays the same but the wavelength of the
perturbation gets shorter ($\propto \beta^{-1/2}$).
    For sufficiently small wavelengths the damping
due to numerical viscosity  ($\sim \nu_{\rm num}k^2$) will be
larger than the MRI growth rate. To generate MRI-driven accretion
in numerical simulations, one should have large enough $\beta$ in
the disc so as to have instability, and at the same time high
enough grid resolution to avoid the damping of small wavelengths.
This theoretical analysis helps in understanding the initial stages of
the MRI instability in our simulations.

\subsection{Numerical setup}
\label{sec:method}

\subsubsection{Method and grid}
\label{sec:numerical-method}

We use three-dimensional (3D) MHD second-order Godunov-type
code developed by our group \citep{kold02}.   It has many
specific features which are oriented towards efficient calculation
of accretion on to a star with a tilted dipole or more complex
magnetic fields: (1) the magnetic field $\bf B$ is decomposed into
the ``main'' dipole component of the star, ${\bf B}_0$, and the
component ${\bf B}_1$ induced by currents in the disc and the
corona \citep{tana94}; (2) the MHD equations are written in a
reference frame which is rotating with the star; and (3) the
numerical method uses the ``cubed sphere" grid which has the
advantages of both the cartesian and spherical coordinate systems
and avoids the singularity on the polar axis in the spherical
coordinate system. The grid on the surface of the sphere consists
of six sectors with the grid in each sector topologically
equivalent to the grid on a face of a cube (e.g.,
\citealt{ronc96}). In contrast with these authors, we use a
Godunov-type numerical scheme similar to the one described by
Powell et al. (1999) and perform simulations in the
range of azimuthal angles ($0-2\pi$) in
three dimensional space.

In addition, to better resolve the MRI-driven turbulence
inside the disc, we compressed the grid towards the equatorial
plane (see Fig. \ref{cubed-comp}). This helps to increase the grid
resolution in the vertical direction, which is favorable for the
investigation of MRI-driven accretion and helps to save computing
time. A number of angular grid resolutions has been used:
$N_x=N_y=51, 61,
 71, 81, 91$ (in each of the 6 blocks of the
cube).  MRI-driven turbulence has been observed in all
 these cases, and the picture of the accretion flow is
  qualitatively the same.
       To save computing time, we use the grid
 $N_x=N_y=61$ as a base.
         The number of grid cells
  in the radial direction is
$N_r=200$ (in the strong field case) and $N_r=220$ (in the weak
field case), so that the region is 1.6 times larger in the case of
a weak field.
      The grid compression has been done in such a way
that the number of grid cells across the disc in the vertical
direction is $80$, which is sufficient for resolving the MRI-driven turbulence. The 3D MHD
code is parallelized using MPI. We typically use $N_p=240-360$
processors per run.

\subsubsection{Reference Units}
\label{sec:refunits}

3D MHD equations are written in dimensionless form and results can be
applicable to a wide variety of stars.  We chose mass of a star as a reference unit
for mass, $M_0=M_\star$ and radius of the star as reference unit for length,
$R_0=R_\star$. \footnote{The magnetospheric radius, $r_m$ is another important scale
of the problem, which is one of the main parameters used in theoretical
investigations. However, in simulations it varies in time, and hence we use radius of
the star as a reference unit, and obtain $r_m$ from simulations.} Reference value for
velocity is Keplerian velocity at radius $R_0$: $v_0=(GM_0/R_0)^{1/2}$. The reference
time scale is period of rotation at $R_0$:  $P_0=2\pi R_0/v_0$.
From dimensionalization of equations, we get ratio:
$\rho_0 v_0^2=B_0^2$, where $B_0$ and $\rho_0$ are the reference magnetic field and
density at $R_0$. We take the reference magnetic field $B_0$ such that the reference
density is typical for considered types of stars (see Tab. \ref{tab:refval}).
 We then define the reference dipole moment $\mu_{0}=B_0R_0^3$,
density $\rho_0=B_0^2/v_0^2$,
mass accretion rate $\dot{M}_0=\rho_0v_0R_0^2$,
 energy per unit time
$\dot{E}_0=\rho_0v_0^3R_0^2$.  Temperature
$T_0=p_0/(\mathcal{R}\rho_0)$, where $\mathcal{R}$ is the gas
constant.

We fix these reference
units (and hence, the initial density, pressure, etc. distribution in the disc), but vary the initial
magnetic field on the surface of
the star and in the disc with dimensionless parameters $B'_\star$
and $B'_d$ respectively \footnote{Note that parameter $B'_\star$
in this paper is identical to parameter $\widetilde\mu$ used in our previous papers (e.g., in
\citealt{roma11a}).}, so that the dimensional magnetic fields on the
surface of the star (at the equator) is: $B_\star=B'_\star B_0$
and initial field in the disc $B_d=B'_d B_0$.

The MHD equations are solved using dimensionless
variables $\widetilde{r}=r/R_0$, $\widetilde{v}=v/v_0$,
$\widetilde{t}=t/P_0$, $\widetilde{B}=B/B_0$,  and so on. In the
subsequent sections, we show dimensionless values for all
quantities and drop the tildes. Our dimensionless simulations are
applicable to stars of different scales. We list the reference
values for a few types of stars in Tab. \ref{tab:refval}. To
derive the real, dimensional values, one needs to multiply the
dimensionless values obtained from simulations by reference values
presented in Tab. \ref{tab:refval}.

\begin{table}
\centering
\begin{tabular}{l@{\extracolsep{0.2em}}l@{}lll}

\hline
& CTTSs       & White dwarfs          & Neutron stars           \\
\hline

{$M_\star(M_\odot)$}              & 0.8              & 1                     & 1.4                     \\
{$R_\star$}                       & $2R_\odot$       & 5000 km               & 10 km                   \\
{$R_0$ (cm)}                      & $1.4\e{11}$      & $5.0\e8 $             & $1.0\e6$                \\
{$v_0$ (cm s$^{-1}$)}             & $2.8\e7$         & $5.2\e8$              & $1.4\e{10}$                \\
{$T_0$ (K)}                       & $9.4\e6$         & $3.2\e9$              & $2.2\e{12}$             \\
{$P_0$}                           & $0.37$ days      & 6.1 s                 & 0.46 ms                  \\
{$B_{\star 0}^{eq}$ (G)}          & $5\e2$           & $5.0\e5$              & $5\e8$                  \\
{$B_0$ (G)}                       & 50               & $5.0\e4$              & $5.0\e7$                \\
{$\rho_0$ (g cm$^{-3}$)}          & $3.3\e{-12}$     & $9.4\e{-9}$           & $1.3\e{-5}$             \\
{$\dot M_0(M_\odot$yr$^{-1})$}    & $5.6\e{-8}$      & $1.9\e{-8}$           & $2.9\e{-9}$             \\
{$\dot E_0$ (erg s$^{-1}$)}       & $1.3\e{33}$      & $3.2\e{35}$           & $3.4\e{37}$             \\
\hline
\end{tabular}
\caption{Sample reference values for typical CTTSs, white dwarfs,
and neutron stars. The dimensional values  can be
obtained by multiplying the dimensionless values by
these reference cvalues. } \label{tab:refval}
\end{table}

\begin{figure}
\centering
\includegraphics[width=8.0cm]{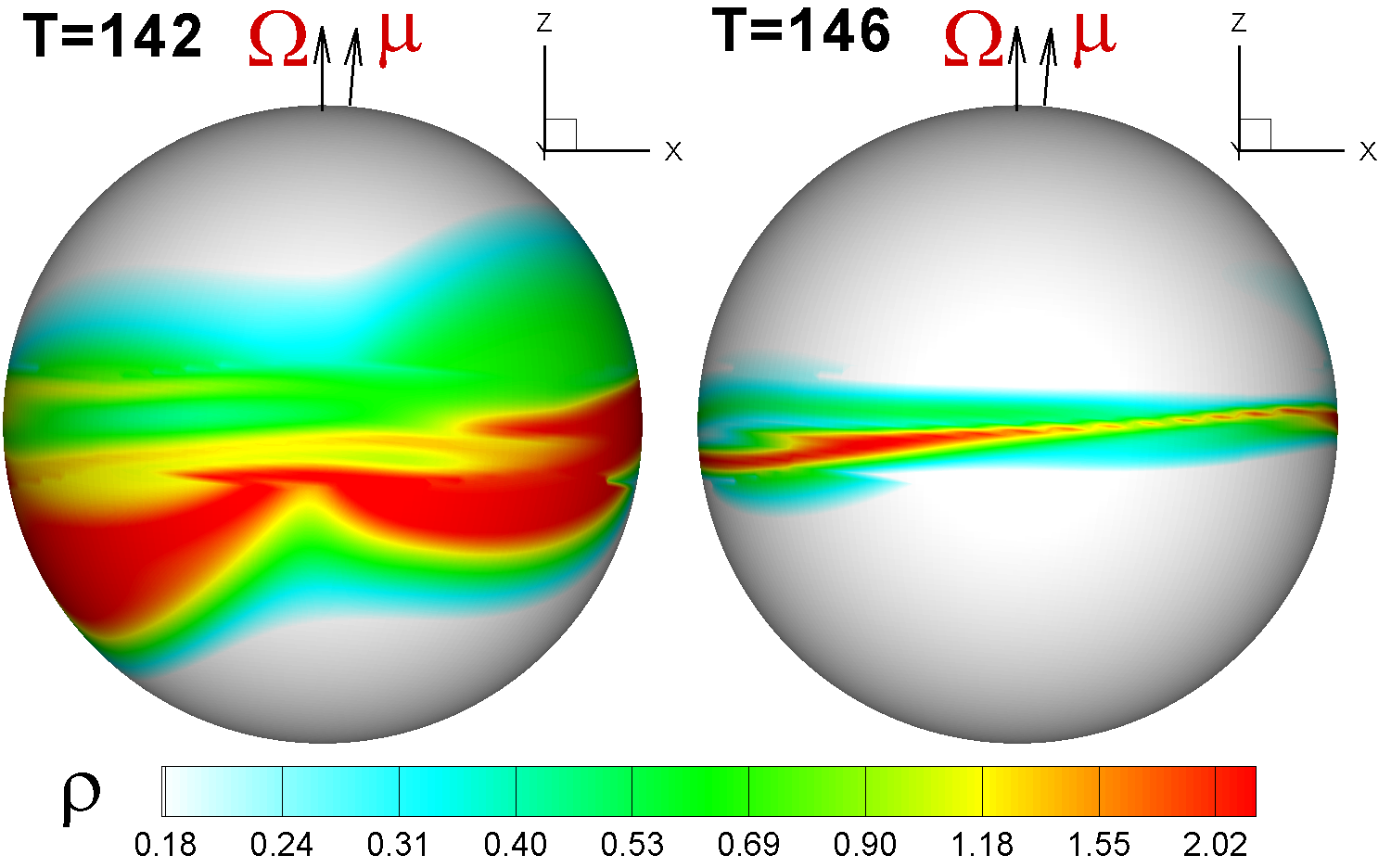}
\caption{Density distribution in the hot spots
at the surface of the star
in the BL regime ( $B'_\star=0.001$)
during the local maximum of the accretion rate, $T=142$ (left panel)
and local minimum $T=146$ (right panel).}
\label{spots-bl}
\end{figure}

\begin{figure}
\centering
\includegraphics[width=8.5cm]{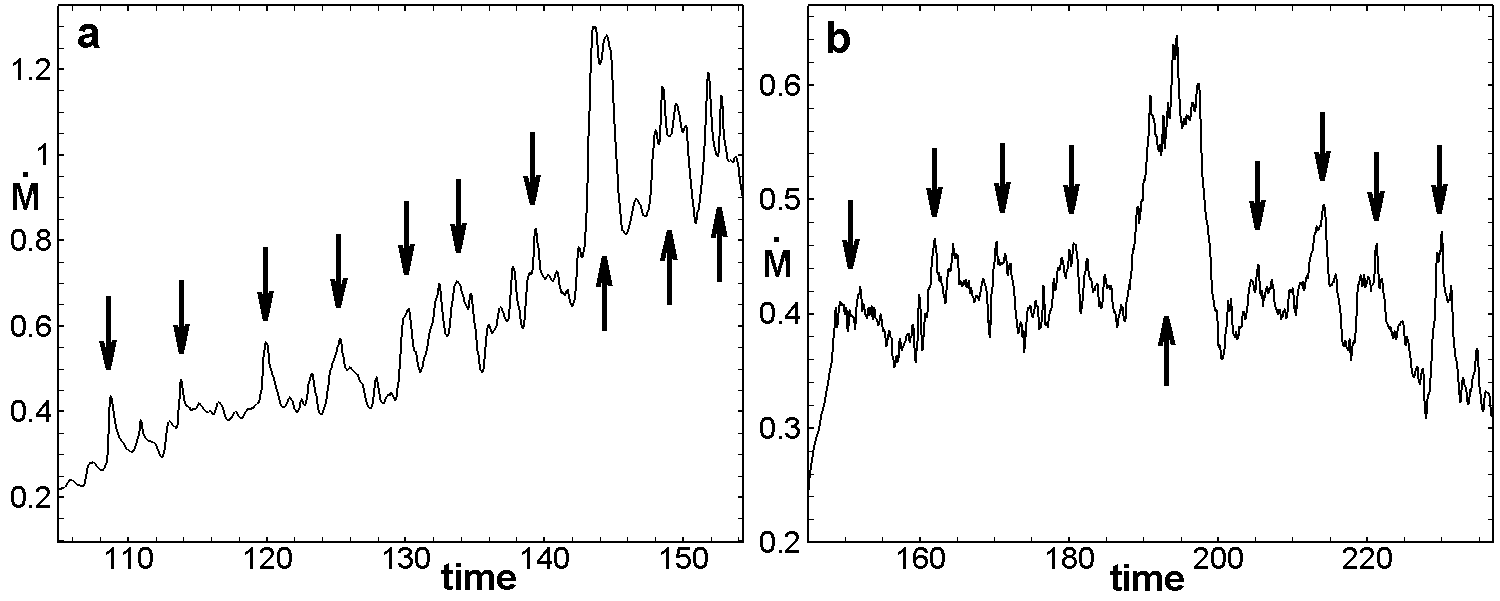}
\caption{Accretion rate at the surface of the star in cases of
the BL regime. Panels a and b show the cases of
$B'_\star=0.001$ and $B'_\star=1$ respectively.} \label{mdot-2}
\end{figure}

\subsubsection{Initial conditions} \label{sec:init}

A rotating magnetized star of mass $M_\star$ and radius $R_\star$ is
surrounded by an accretion disc and a corona. The disc is relatively cold and dense,
while the corona is hot and rarefied. We set the temperature in the corona $T_c=T_0$
and in the disc $T_d=0.01 T_c$ (we determine these values at the inner edge of the
disc, which is our reference point). The density in the disc and corona
$\rho_d=\rho_0$, $\rho_c=0.01\rho_d$. Here, the subscripts `d' and `c' denote the disc
and the corona. The initial density distribution is derived from the balance between
the gravitational, centrifugal and pressure gradient forces, as well as the condition
that the corona rotates with the angular velocity of the disc and the gas is initially
barotropic. In addition, the pressure is balanced at the boundary between the disc and
corona. The resulting initial distributions of density in the disc and corona is
almost homogeneous
 (within 10\%). The density only slightly increases outward in both, disc,
and corona. Distribution of the temperature (and the sound speed $c_s$)
is also almost homogeneous. In
dimensionless units, initial values at the reference point are:
$\rho_d=1$, $\rho_c=0.01$, $T_d=0.01$, $T_c=1$.

The star rotates slowly with a fixed angular velocity
$\Omega_\star$ such that the corotation radius
$r_{\rm cor}=10 R_\star$, so that $r_{\rm cor}/r_m\approx 3$.
We take such a slow rotation to be sure that
conditions are favorable for magnetospheric accretion during the
whole simulation run, as we discussed in Sec. \ref{sec:setup-rotation}.
Cases of more rapid rotation and the propeller regime will be
investigated in the future.

\begin{figure*}
\centering
\includegraphics[width=17.0cm]{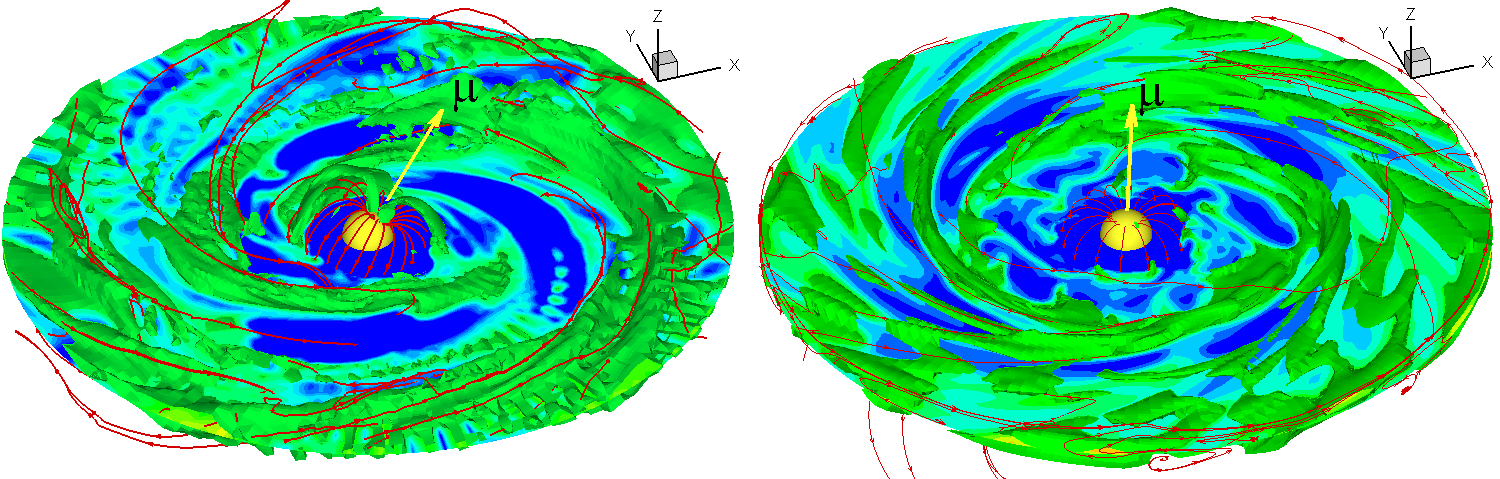}
\caption{A three-dimensional view of accretion on to a star with a
large magnetic field, $B'_\star=10$, at large
$\Theta=30^\circ$ (left-hand panel) and small $\Theta=2^\circ$
tilts.  The green background shows one of density levels, the blue
color shows the equatorial density slice, and the lines are sample
magnetic field lines.} \label{3dmri-t2-t30}
\end{figure*}

A star has a dipole magnetic field. The strength at the equator is
determined by the dimensionless parameter  $B'_\star$.
The dipole moment is tilted at an angle $\Theta$ about the rotational axis of the star
(which is aligned with the rotational axis of the disc).
Simulations show that at typical parameters of the disc, described above,
the disc is truncated by the magnetosphere (and hence, the magnetosphere is dynamically important) if
$B'_\star \gtrsim  1$. At smaller values of the field, the magnetosphere is dynamically unimportant,
and matter accretes in the BL regime. We take very weak field, $B'_\star=0.001$,
for modelling the BL regime. We also take the marginal field, $B'_\star=1$, for test runs.
A stronger field, $B'_\star=10$, is used for modelling the magnetospheric
regime of accretion.

The disc and corona are threaded with a homogeneous small seed poloidal field.
The strength of the field is determined by the dimensionless
parameter $B'_d$. We take $B'_d=0.005$ in case of the BL
regime, and $B'_d=0.01$ in case of the magnetospheric regime.
Figure \ref{init-3} shows the initial distribution of density
and magnetic field in $xz-$slices in cases of the BL regime, where $B'_\star=0.001$, $\Theta=5^\circ$ (left panel),
and magnetospheric regime, where  $B'_\star=10$, and the tilt
is either large, $\Theta=30^\circ$ (middle panel), or very small,
$\Theta=2^\circ$ (right panel).

Conversion of our dimensionless parameters to dimensional shows that the
density, velocity, magnetic field and other parameters match well with those observed
in typical stars listed in Tab. \ref{tab:refval}. However, the temperature in both,
disc and corona, is about 10-20 times higher than that observed in real stars. In our
initial setup, this is a consequence of the low density contrast between the disc and
corona, which is initially  $\rho_c/\rho_d=10^{-2}$. At the higher density contrast,
$\rho_c/\rho_d=10^{-3}$, temperature is realistic, however these simulations require
higher grid resolution and the simulation time is much longer. Realistic, much lower
values for temperature in the disc and corona are used in \textit{axisymmetric}
simulations of $\alpha-$discs (e.g., \citealt{long05}) and MRI-driven discs
\citep{roma11a}. These simulations do not show any notable difference in results
between high and low-temperature flows. We suggest that in both cases, the inner disc
is sufficiently thick, so that the gravity force is the main force  pulling matter to
the funnel flow (if $r_m < r_{\rm cor}$). We discuss possible influence of the higher
temperature in the disc to magnetospheric accretion in Sec.
\ref{sec:disc-magnetosphere}.

\begin{figure*}
\centering
\includegraphics[width=14cm]{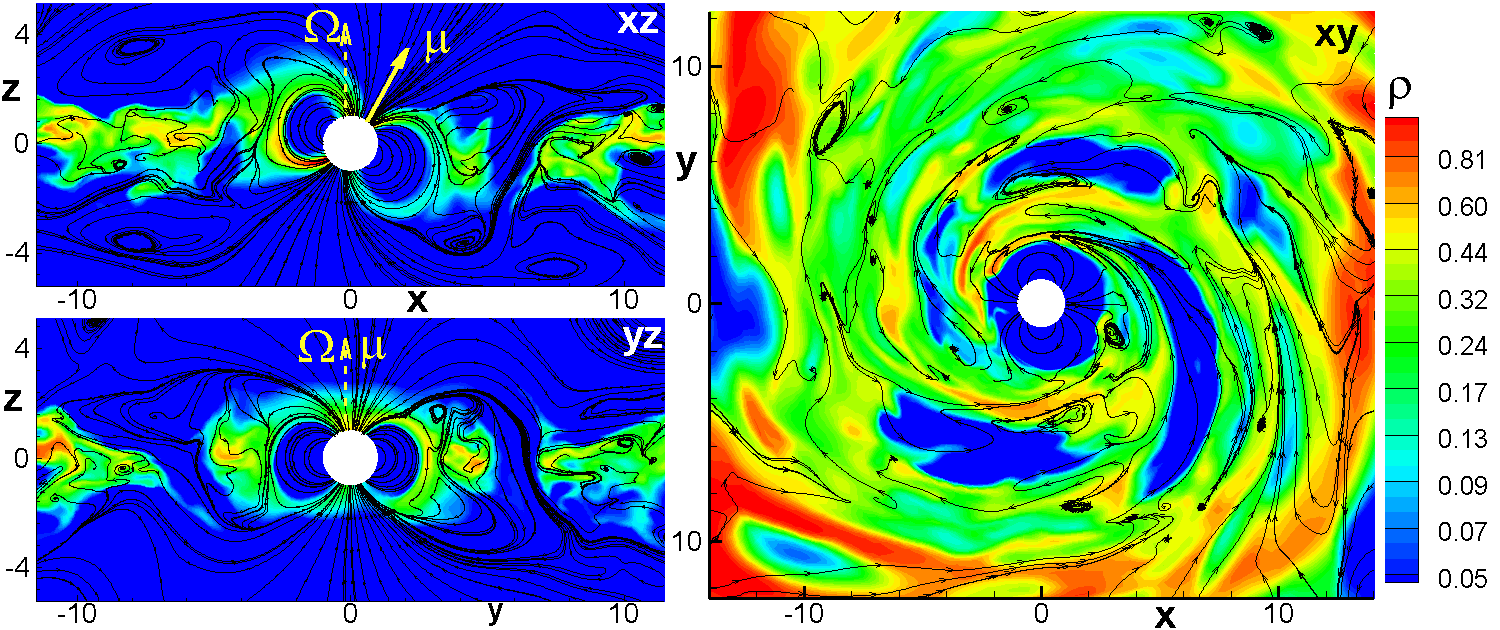}
\caption{
Slices of the density distribution (color background) and
sample field lines in the case of a star with dynamically
important magnetic field ($B'_d=10$) and large tilt ($\Theta=30^\circ$). The
left-hand panels show the $xz$ and $yz$ slices, while the
right-hand panel shows the equatorial slice, $xy$,
}
\label{d10-t30-slice-3}
\end{figure*}

\subsubsection{Boundary conditions} \label{sec:bound}

At the outer boundary of the simulation region, we fixed
all 8 variables. The boundary is located at $R_{\rm
out}\approx 51 R_\star$  for the case of the BL
regime and $R_{\rm out}\approx 32 R_\star$ for the magnetospheric regime. We observed from
simulations that the disc is sufficiently large to supply matter
during the whole simulation run.

At the inner boundary (which is a stellar surface) we take  `free'
 boundary conditions, which are corrected to the specifics of the considered
 problem. In general, the number of the boundary conditions should correspond
 to a number of waves propagating from the boundary to the simulation region.
However, the relationship between the radial velocity at the
inner boundary and velocities of MHD waves is not known in advance. In addition,
the interaction of the accreting matter with the stellar surface is a separate,
complex problem, which requires detailed calculations of processes at much smaller
 scale \footnote{For example, interaction of the supersonic funnel stream with
 stellar surface leads to the formation of the radiative shock wave, which
 oscillates (e.g., \citealt{kold08}).}.
Besides, matter   can accrete, or,
it can flow away from the star.
 We do not consider the complex
processes near the stellar surface, but instead use simplified boundary
conditions, which allow the absorption of matter which falls to the star's
surface.

We consider two types of boundary conditions
for different regimes of accretion.  \textit{Type A:} If the
magnetic field of the star is dynamically unimportant
and accretion is super-Alfv\'enic, then we use `free' boundary condition
 for all variables: $\partial U/\partial r = 0$ where $U = s$
 (entropy), ${\bf v}$ (total velocity vector), ${\bf B}_t$ (tangential to
 the stellar surface component of the field), $r^2 B_r$ (normal to the
 surface component of the field multiplied by $r^2$). We artificially decrease the pressure $p$
 at the inner boundary so that to stimulate accretion, and to remove
  excessive matter if it accumulates at the boundary:
$p_0 = k p_1,~~k=0.9$, where the subscripts $'0'$ and $'1'$ mark the boundary grid and
the last calculated cell respectively. The density in the boundary cell is
recalculated as $\rho_0 = (p_0/s_0)^{1/\gamma}$. These conditions are used in the
regime of the boundary layer accretion.

\textit{Type B:}
 In case when the magnetic field is dynamically important, then the accreting flow is sub-Alfv\'enic
 (at the surface of the star) and two waves propagate from the boundary
 to the simulation region: Alfv\'en wave and fast magnetosonic wave.
  Hence, we provide two additional boundary conditions: in the coordinate
   system rotating with the star, the velocity vector is parallel to the
    magnetic field vector, which we split to two components:
    normal (or, radial) component, $B_r$, and the component tangential to the surface of the star, $B_t$.
     This condition means that the strong magnetic field acts as `rails'
      attached to the non-deformed surface of the star along which matter
      flows to the star. We use the following algorithm to calculate these
       conditions:
1) we calculate the total vector of velocity in the boundary grid
as ${\bf v}_0^{'} = {\bf v}_1$;~
2) we subtract rotation of the star:
${\bf v}_0^{''} = {\bf v}_0^{'} - (\Omega_\star \times {\bf r}_0 )$,
where ${\bf r}_0$ is the radius-vector
of the center of the boundary grid,~
3) we project vector ${\bf v}_0^{''}$ to the direction of the magnetic
 field: ${\bf v}_0^{'''} = ({\bf B}_0{\bf v}_0^{''})/{B_0^2}$;
4) we add to the velocity a part connected with rotation of the star:
${\bf v}_0 = {\bf v}_0^{'''} + (\Omega_\star \times {\bf r}_0 )$.
We use these boundary conditions for modelling the magnetospheric regime.

\subsection{Number of MRI wavelengths per thickness of the disc}
\label{sec:number-modes}

Before starting simulations, we estimate the expected number of MRI wavelengths within the
thickness of the disc, using the analysis outlined in
Sec. \ref{sec:setup-mri-theory} and the dimensionless parameters of the model.
The wavelength of the most unstable,
MRI-driven mode is $\lambda_{\rm MRI} \approx 2 \pi
{v_{A,z}}/{\Omega_K}$. In the approximation of the thin
disc, the full thickness of the disc is $2 h \approx {2
c_s}/{\Omega_K}$, where $c_s = \sqrt{p/\rho}$ is the locally isothermal
sound speed in the disc. The number of wavelengths per
thickness of the disc is $N_{\rm MRI} = {2 h}/{\lambda_{\rm MRI}}
\approx {c_s}/{\pi v_{A,z}}$.
 In our model
 (in dimensionless units): density in the disc is $\rho_d \approx 1$, the sound speed is $c_s \approx 0.1$,
the Alfv\'en velocity (based on $B'_d=0.01$ field) is $v_{A,z}
= {B'_d}/{\sqrt{4 \pi \rho_d}} \approx 2.8\times 10^{-3} (B'_d/0.01)(\rho_d/1.0)^{-1/2}$.
Substituting these values to the initial formula for $N_{\rm MRI}$ and to eq. \ref{eq:beta-theor}, we obtain:

\begin{equation}
\beta = {2 c_s^2 \over v_A^2} = 2.5\times 10^{3}
\bigg(\frac{0.01}{B'_d}\bigg)^2 \bigg(\frac{\rho_d}{1.0}\bigg) \bigg(\frac{c_s}{0.1}\bigg)^2 ~, \label{eq:beta-theor}
\end{equation}
\begin{equation}
N_{\rm MRI} = \frac{2 h}{\lambda_{\rm MRI}} = 11
\bigg(\frac{0.01}{B'_d}\bigg)
\bigg(\frac{\rho_d}{1.0}\bigg)^{-1/2} \bigg(\frac{c_s}{0.1}\bigg)~.
\label{eq:N-mri}
\end{equation}
These formulae are approximate  but useful for understanding the start-up
conditions for development of the instability.
Below, we show results of our simulations
in two different regimes: the BL regime, and regime of the
magnetospheric accretion.

\begin{figure*}
\centering
\includegraphics[width=14.0cm]{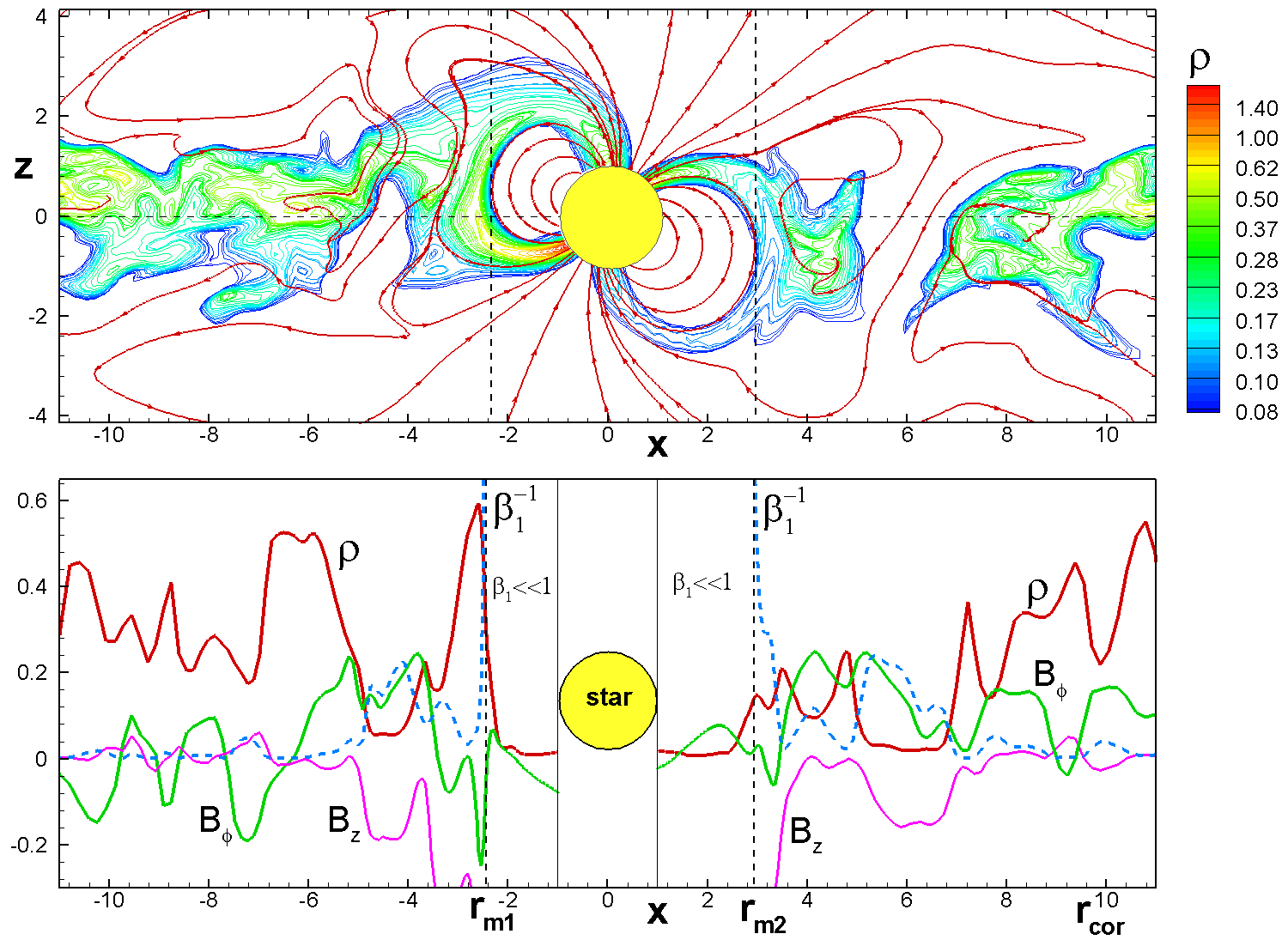}
\caption{{\it Top panel:} The density distribution in the inner disc and in funnel
streams in the $xz-$plane and selected field lines. {\it Bottom
panel:} Linear distributions of different variables in x-direction
in the equatorial plane: the density,$\rho$, two components of magnetic field,
$B_z$ and $B_\phi$, and $\beta_1^{-1}$. The dashed vertical
lines correspond to $\beta_1=1$ and show the position
of the magnetospheric radius to the left, $r_{m1}$ and to the right, $r_{m2}$,
of the star.   The position of the corotation radius,  $r_{\rm cor}$, is also shown.}
\label{fun-lin-2}
\end{figure*}

\section{Boundary layer regime}
\label{sec:mri-mu001}

First, we investigate the case where the magnetic field of
the star is dynamically unimportant and matter accretes to the star in the BL regime.
We use this case to investigate the MRI-driven turbulence in the disc in our
numerical setup. This case is also interesting from the
astrophysical point of view, because many stars are expected to accrete
in the BL regime. Below, we discuss both aspects of the problem.

\subsection{MRI-driven accretion}

We investigated accretion at several values of the seed poloidal field in the disc
from $B'_d=0.002$ up to $B'_d=0.01$, and
found that in all cases the MRI-driven turbulence is developed, matter accretes inward,
and angular momentum is transported outward by the magnetic shear stress
(see details is Sec. \ref{sec:stresses}).
 In experiments with the smallest in the set magnetic field, $B'_d=0.002$ \footnote{We used this field
 as a base in
 axisymmetric simulations \citep{roma11a}.}, the initial number of
wavelengths per thickness of the disc is relatively large, $N_{\rm MRI}=55$ (at the inner disc),
 but the accretion rate, which is proportional to the magnetic
stress in the disc, is low: matter of the inner disc reaches the star in $T\approx 200$ rotations (at $r=1$).
 These simulations last up to $T\approx 800$.
     In the case of the largest field, $B'_d=0.01$ ($N_{\rm
MRI}=11$), the accretion rate is high, however only 1-2 turbulent
cells per thickness of the disc are observed in developed
turbulence. \footnote{Note, that the initial value of $N_{\rm MRI}$ is based
on the \textit{initial} seed poloidal field in the disc. Subsequently, the
azimuthal component increases, and the number of
turbulent cells per thickness of the disc drops.} We take
the intermediate field, $B'_d=0.005$, as a base for simulations,
because the accretion rate is relatively high, and at the
same time, a few turbulent cells per thickness of the disc are
observed.

We also varied the magnetic field of the star from very
weak, $B'_\star=0.001$ (and dynamically insignificant during the
whole simulation run),  to somewhat stronger,
$B'_\star=1$, where a tiny magnetosphere was observed initially,
but it disappeared later, when the accretion
rate become larger.

The initial density and temperature distribution in the disc is almost
homogeneous (within $10\%$), and the magnetic field threading the disc and corona is a
constant (see Fig. \ref{init-3}, left panel, and Sec. \ref{sec:init}).
      As a result, the initial distribution of $\beta-$parameter in the disc
 is almost homogeneous, and for the
disc field of $B'_d=0.005$, it is $\beta\approx 9,000$. We
observed, that the initial number of wavelengths per thickness of
the disc is $N_{\rm MRI}\approx 10-20$, as predicted by the theory (see
Sec. \ref{sec:number-modes}).
     Subsequently, the  perturbations
grow due to the magneto-rotational instability,
and  turbulence  develops in the
inner parts of the disc (at $T\approx 10-20$ periods of rotation at
$r=1$).  Later, turbulence develops at larger radii. Subsequently,
the magnetic energy in the disc increases, mainly due to the growth of
the azimuthal component of the field, which leads to strong
decrease of $\beta$. As a result, the number of turbulent cells
per thickness of the disc decreases up to 2-3.

Figure \ref{d001-t5-slice-3-vac} shows slices of the density
distribution at $T=150$. The left-hand panels show that the flow
is turbulent, and the poloidal field lines are tangled and have a
number of reversals which track different turbulent cells. The
right-hand panel of Fig. \ref{d001-t5-slice-3-vac} shows the
equatorial slice of the density distribution. One can see that the
turbulent cells are strongly elongated in the azimuthal direction,
and they also look like parts of spiral waves.

\subsection{Azimuthally-wrapped magnetic field in the inner disc}

In the inner parts of the disc, the azimuthal magnetic field
increases rapidly due to strong shear in the flow.
In addition, the poloidal
magnetic field increases towards the inner disc
due to the convergence of the flow and conservation of the poloidal magnetic flux.
As a result, the azimuthal component of the field reaches the value
 of $B_\phi\approx 1$ which is about 200 times larger than the initial seed poloidal
 field of $B'_d=0.005$.
 The plasma parameter  $\beta$ drops to $\beta\approx 1-3$,
 that is, the magnetic and thermal pressure
 are close to the equipartition state.

Fig. \ref{d001-t5-beta-2} (right panel) shows the distribution of the magnetic pressure
scaled to matter pressure,  $\beta^{-1}=B^2/8\pi p$ at $t=150$.  The thick black line
shows exact
equipartition, where $\beta=1$. One can see elongated islands,
 where the magnetic pressure dominates (yellow and
red color). The left-hand panel shows the $xz-$slice, where the
turbulent cells are seen in their cross-sections.
 These simulations are in accord with
 earlier simulations by  \citet{armi02} and  by \citet{stei02},
 who also observed strong amplification of the field
 in the inner disc.

\citet{prin89} argued
that  magnetic flux built up in the disc should escape
 to the corona due to the Parker instability, and that this field can be responsible for magnetically-driven
outflows from young stars (see also \citealt{shu88}).
        Formation of the magnetically-dominated corona
has been observed in 3D simulations (e.g.,  \citealt{mill00,hawl02}).
Very long-lasting axisymmetric simulations by \citet{roma11a} performed at 10 times lower density
in the corona (compared with present simulations) show the formation of a very large magnetic corona
which slowly moves away from the disc to large distances, and which is driven by the magnetic force
 (see fig. 20 and 21 of \citealt{roma11a}).
    In the present simulations, we have not seen such a corona, because the matter density in the corona
is too high, and simulations are not as long. We expect that the large-scale expanding
magnetic corona will be seen in the future 3D simulations. Formation of faster outflows from the disc-star,
boundary, as discussed by \citet{prin89}, also can no be excluded.

We also observed an interesting phenomenon, that
the magnetic flux of the star
increases in time, and
a new  magnetosphere forms around the star.
The initially weak magnetic field of $B'_\star=0.001$ increased
up to $B_\star\approx 1$ (at the pole, above the star) because the accreting matter
brings the poloidal field towards the star, and this field is accumulated at the
stellar surface. Fig. \ref{d001-t5-beta-2} (left panel) shows that the new magnetosphere
has a monopole-type shape. The magnetosphere does not truncate the disc
\footnote{Note that the field $B'_\star\approx 1$ corresponds to the border
 between dynamically active
and passive magnetospheres.},
 and the disc accretes in the BL regime.
We suggest that in longer simulation runs, the magnetic field can increase up to even larger values. On the other hand, the disc can bring matter with the field of one or another polarity, and the induced magnetic field of the star will flip its polarity.

Of, course, details of the disc-star interaction depend on the boundary conditions at the star.
Here, we show results for one type
of boundary conditions (\textit{Type A}, see Sec. \ref{sec:bound}), where both, matter and magnetic field
can accrete on to the surface of the star.
However, conditions at the boundary can be different. For example, if
a star is a perfect conductor,
then the magnetic flux of the disc will not penetrate through the star, but instead it will
accumulate at the disc-star boundary. Such a condition will be more favorable for formation of the magnetic torus, and/or for outflows.

\begin{figure}
\centering
\includegraphics[width=8.0cm]{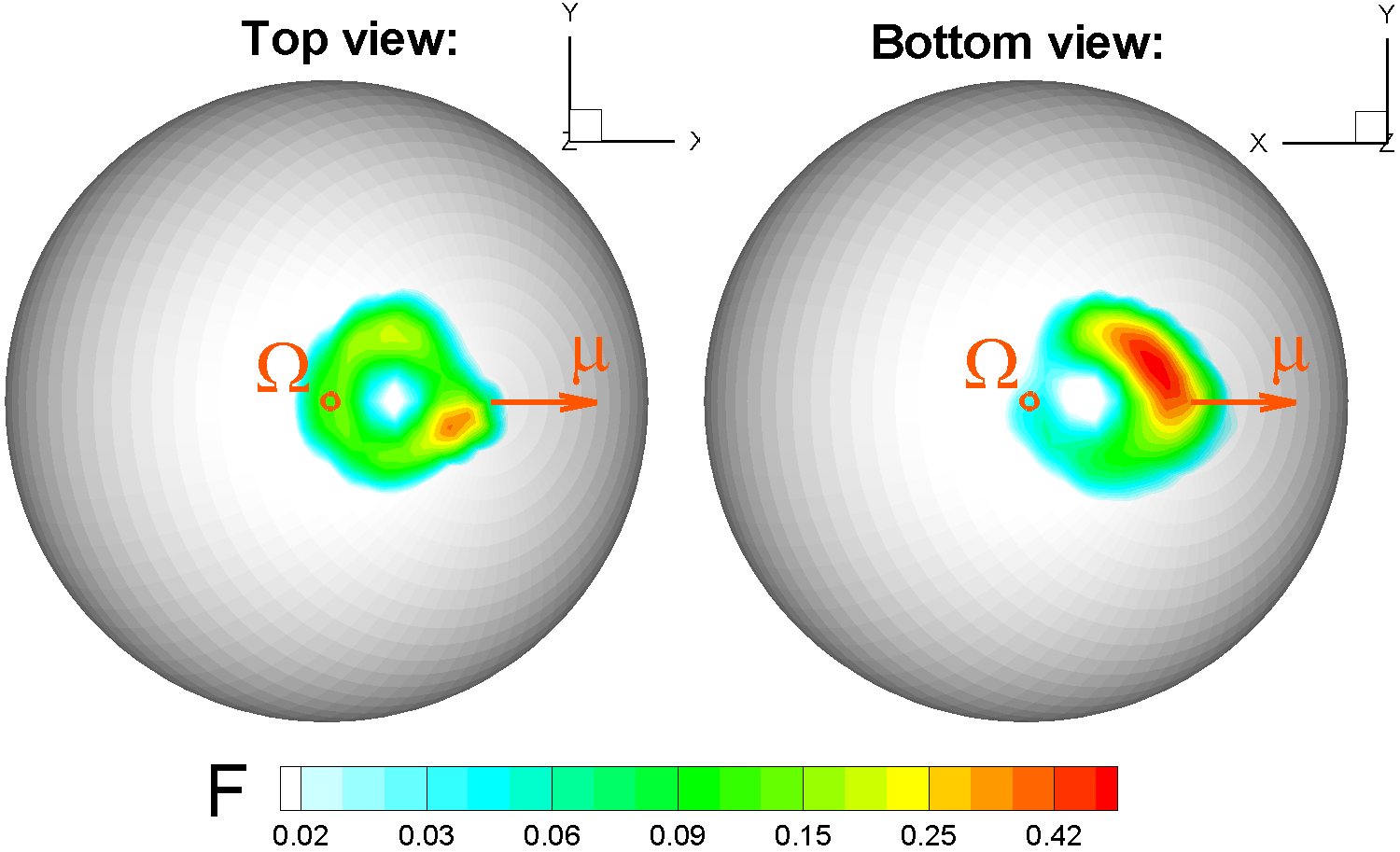}
\caption{Distribution of the kinetic energy flux in the hot spots
at the surface of the star as seen from the top ($-z$-direction,
left-hand panel) and the bottom ($+z$-direction, right-hand
panel).}\label{spots-2}
\end{figure}

\subsection{Equatorial belt spots and variability associated with accretion of turbulent cells}

In the BL regime, the disc matter accretes to the star in the region of the equator, and it
forms the belt-shaped hot spot.
About a half of the gravitational energy of the accreting matter is expected to be released in the equatorial belt spot.
The details of the physics of the disc-star interaction including the processes of the heating and cooling of matter in the BL are not
established.
The  interaction may include complex processes such as the Kelvin-Helmholtz instability and the formation of
a turbulent layer at the surface of the star,.
One can expect, however, that most of energy will be released in regions
of enhanced density. This is why we show the density distribution at the star's surface.

We observed in simulations that matter accretes onto the star in azimuthally-stretched turbulent cells.
Near the star, the turbulent cell becomes stretched in the azimuthal direction, and when it accretes on to the star, it forms
the higher-density spot which is also spread in the meridional direction. The spots are less dense and more narrow during periods of lower accretion rate.

Fig. \ref{spots-bl} (left panel) shows the density spot during accretion of one of the turbulent cells.
The spot is not symmetric,  because the turbulent cell is not azimuthally-symmetric. The spot is quite wide in the meridional direction.
The right panel shows the spot during period of the minimum of the accretion rate, when one of cells already accreted, while another one did not approach the star yet.
This spot is more narrow in the meridional direction.

Accretion of turbulent cells  leads to \textit{variability} of the accretion rate at the star.
Fig. \ref{mdot-2} shows the accretion rate for our reference case with initial stellar field of
$B'_\star=0.001$ (left panel), and in the case of larger initial
field of the star, $B'_\star=1$ (right panel).
   For these cases, we used boundary conditions of \textit{Type A} and \textit{Type B}
respectively (see Sec. \ref{sec:bound}).
     We observed that the peaks
in the accretion rate (see arrows in the plot) correspond to the accretion of  individual turbulent cells.
    The accretion rate is typically about   20\% higher than the average during accretion of a turbulent
cell, and the maxima at 50\%
correspond to the accretion of the largest turbulent cells.
    The time interval between
peaks in these two cases is $\Delta t\approx 5$ and
$\Delta t\approx 10$, respectively.
     The accretion rate curves are quite similar in spite of the
different initial fields of the star and somewhat different boundary conditions.
   Note that we have the same initial conditions in discs,
and hence the MRI-induced turbulence in the disc has the same properties. This may determine the similarity in the time-scale
between peaks. Some difference is connected with different conditions at the star.

In our simulations, we observe
 quite large  turbulent cells, and hence we observe
and `count' them during their accretion on to the star.  Such turbulence dominates in
discs  which are close to equipartition with $\beta \sim 1$. We suggest, that
equipartition is a probable  state in the inner regions of accretion discs where the
shear is high. In the opposite case, $\beta>>1$,  multiple turbulent cells are present
in the disc. The simultaneous accretion of multiple cells will lead to much smoother
accretion, and results will be closer to those observed in cases of $\alpha-$discs
(e.g., \citealt{roma04}). Hence, the variability pattern carries information about the
 property of the turbulence in the disc.

\section{Regime of Magnetospheric Accretion } \label{sec:mri-mu10}

Next, we discuss  accretion on to stars with \textit{dynamically
important} magnetic fields, where the disc is truncated by the
magnetosphere of the star and matter accretes in funnel streams.
We chose the parameter $B'_\star=10$ for these runs.
     We investigate cases of large ($\Theta=30^\circ$) and small
($\Theta=2^\circ$) tilts of the dipole magnetic moment relative to
the rotational axis.
      To save computing time, we increased the seed
magnetic field in the disc up to $B'_d=0.01$. This helped to
increase the accretion rate (which is proportional to the
magnetic stress in the disc), and cause the disc to reach the
surface of the star more rapidly, at about
$T=40$\footnote{ Simulations with a field of $B'_d=0.005$ produce
similar results, but on a longer time scale.}.

We observed that initially, about 10 MRI wavelengths
fit within the vertical extent of the disc, and the inner parts
of the disc started accreting towards the star. Later, the
azimuthal field increased, and the number of turbulent cells observed later in simulations dropped to 1-2 per thickness of the disc.

Figure \ref{3dmri-t2-t30} shows a 3D view of the MRI-driven discs in cases of low and high tilts of the dipole field. The background
shows one of density levels.
   One can see that in both cases, the disc is strongly inhomogeneous:
matter tends to be concentrated in azimuthally-elongated spiral
segments. The magnetic field lines are stretched in the azimuthal direction.    Below, we consider these two cases in greater detail.

\begin{figure*}
\centering
\includegraphics[width=14cm]{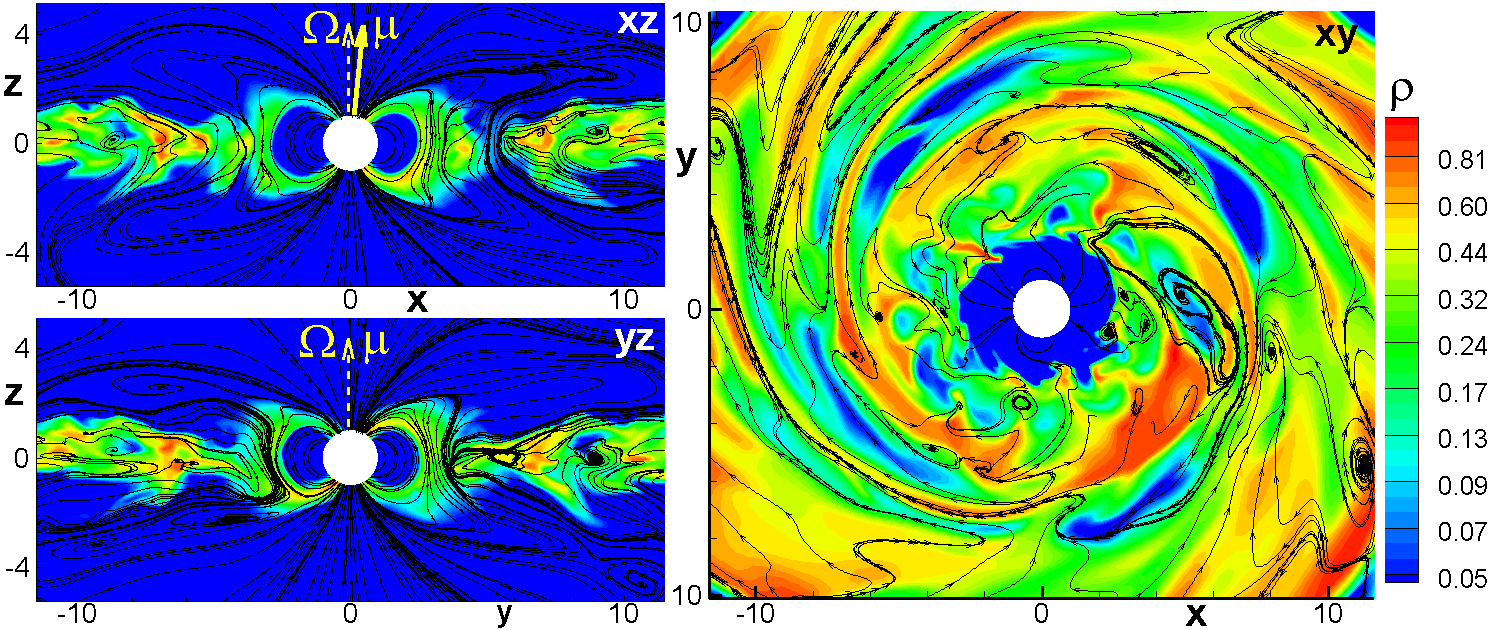}
\caption{Slices of the density distribution (color background) and
sample field lines in the case of a star with dynamically
important magnetic field ($B'_d=10$) and high tilt ($\Theta=2^\circ$). The
left-hand panels show the $xz$ and $yz$ slices, while the
right-hand panel shows an equatorial, slice, $xy$.} \label{d10-t2-slice-3}
\end{figure*}

\subsection{Accretion on to a star with a large tilt, $\Theta=30^\circ$}
\label{sec:disc-magnetosphere}

In the case of a large tilt of the dipole field, matter flows to the star in funnel streams.
Figure \ref{3dmri-t2-t30} (left panel) shows such an accretion
flow, where matter of the disc accretes to the star in funnel streams.
      These funnel streams form when a large turbulent cell moves towards the magnetosphere,
where it is stretched, slowed down by the magnetosphere, and pulled towards the star by the gravity force.

Figure  \ref{d10-t30-slice-3} (left two panels) shows
$x-z$ and $y-z$ slices of density distribution and the poloidal magnetic field.
One can see that the disc is truncated and matter flows to the star in funnel streams.
The height of the magnetosphere is only slightly larger than the height of the turbulent disc.
The regions of enhanced density in the disc (yellow and red colors) track the turbulent cells.
We observed accretion of several turbulent cells during the simulations.

The field lines threading the disc inflate into the corona.
The accretion disc compresses the external field lines and pushes them to reconnect.
After reconnection, the magnetic islands form and propagate into the corona.
This process of compression and reconnection is
similar to that observed in axisymmetric simulations
\citep{roma11a}. Reconnection acts as an efficient means of
diffusivity, and it helps the disc to penetrate through the external layers of the
magnetosphere of the star towards the closed magnetosphere and
to the star through the funnel flow.

The right-hand panel of Fig. \ref{d10-t30-slice-3} shows that the
magnetic field has a significant azimuthal component, and the
turbulent cells are elongated in the azimuthal direction. The
tilted rotating dipole excites two spiral waves in the inner disc.  These waves are similar to
those observed in 3D simulations of $\alpha-$discs with a tilted
dipole magnetic field \citep{roma11b}.

Figure \ref{fun-lin-2} (top panel) shows the disc-magnetosphere boundary in greater detail.
One can see that the disc  is turbulent.
It is disrupted by the stellar magnetosphere and the matter  flows towards the star forming a funnel flow.
One can see that the funnel streams are not symmetric: the accretion rate is higher from the left side
than from the right side.
This is in contrast with earlier 3D simulations of accretion from $\alpha-$discs, where
the funnel streams are symmetric (see, e.g., Fig. 4 from \citealt{roma04}). The asymmetry
is connected with the fact that matter approaches the magnetosphere
of the star and then accretes in large-scale turbulent cells, which are usually closer to one of poles than to another.
Fig. \ref{3dmri-t2-t30} (left panel) shows a typical process of accretion of the turbulent cell.

The bottom panel of Fig. \ref{fun-lin-2} shows the
distribution of different variables in the equatorial
plane. One can see that the density $\rho$ is highly variable in
the inner disc, and it is very small inside the
magnetically-dominated magnetosphere. The azimuthal
component of the field, $B_\phi$, is a few times larger than
$B_z$-component. Inside the magnetosphere, the toroidal
field is much smaller than the poloidal field.

The inner radius of the disc coincides with
the point where our modified plasma parameter
parameter $\beta_1$ (eq. 2) is unity. The vertical dashed lines in Fig. \ref{d10-t30-slice-3}
show the radii where $\beta_1=1$. One can see that the magnetospheric radius
to the left of the star, $r_{m1}\approx 2.4$ is smaller than to the right, $r_{m2}\approx 3$.
The asymmetric nature of the magnetospheric surface is also seen in Fig. \ref{d10-t30-slice-3} (right panel).
Top panel of the Figure also shows that the funnel streams have a finite width
and some matter start flowing to the funnel from larger radii, $r\approx 3-4$.

Matter moves to the funnel flow due to the gravitational force, which dominates over the centrifugal force (in our case
of a slowly rotating star).
  The lifting of matter to $|z|>0$ to the funnel
starts at the radius where the
azimuthal motion of the inner disc matter slows down due
to interaction with the magnetosphere.
    This occurs at $r\approx 4-5$.
     The magnetosphere is small, while the disc is has a finite thickness due to the turbulent  nature of the flow.
      Hence the magnetosphere is not a big obstacle for the flow
 of matter towards the star\footnote{Note, that
the situation is different in cases of much large magnetospheres of X-ray pulsars and strongly magnetized white dwarfs in cataclysmic variables,
where the hight of the magnetosphere may be orders of magnitude larger than the thickness of the disc.}.
   The matter pressure gradient force also contributes to the lifting, though the role of this force is not as significant as in
the case of the \textit{aligned} dipole (e.g., \citealt{roma02} and \citealt{camp10})
Note, that axisymmetric simulations of MRI-driven accretion were performed at much
lower, realistic temperature in the disc and corona, and did not show principal
difference between higher and lower temperature cases  \citep{roma11a}. Moreover, in
case of a tilted dipole, the magnetic field lines are inclined in the equatorial
plane, and hence the star's gravitational force can pull matter directly from the
equatorial plane of the disc.

Note, that inside the disc, the total matter stress is much larger than the magnetic stress, $\beta_1>>1$,
 and the disc acts as matter-dominated (see also Sec. \ref{sec:stresses}).
The MRI-driven instability, initiated by the initial seed field in the disc, provides inward transport of matter, which can be
roughly described as an effective `viscosity'.
However, it does not provide diffusivity at the disc-magnetosphere boundary.
he magnetic field of the star does not thread the incoming disc and does not influence to the MRI processes inside the disc.
As a result the matter-dominated disc compresses the magnetosphere and pushes it to reconnect as shown in Fig. \ref{d10-t30-slice-3}
and Fig. \ref{fun-lin-2}. It also penetrates toward the closed magnetosphere due to instabilities.

The funnel stream hits the surface of the star  with a high
velocity
and the kinetic energy of the flow is released.
    Figure \ref{spots-2} shows  the distribution of the kinetic energy
flux in the hot spots on the surface of the star due to the impact
of the funnel streams. One can see that the spots forming at the
south and north hemispheres of the star are different. The difference
between spots reflects the difference between funnel streams approaching the  south and north hemispheres.
Note, that hot spots obtained  in 3D simulations of
$\alpha-$discs, are almost identical
(\citealt{roma04}). Inequality of spots in the case of turbulent discs may have important consequences for analysis
of variability from accreting magnetized stars.

\subsection{Accretion on to a star with small tilt, $\Theta=2^\circ$ and interchange instability}

Here, we discuss results for the case of very small tilt, $\Theta=2^\circ$.
Figure  \ref{d10-t2-slice-3} (left panels) shows that the magnetosphere
inflates in an almost symmetrical fashion.  The magnetic flux
inflates sideways, as we observed in axisymmetric simulations
\citep{roma11a}. This is because the time scale of disc accretion
is smaller than the inflation time scale. The magnetic field in
the inflated and stretched corona reconnects, while magnetic
islands form and propagate outward.

The right-hand panel of Fig. \ref{d10-t2-slice-3} shows an equatorial slice. One can
see that the magnetosphere with such a small tilt does not excite strong spiral waves.
However, we observed another interesting phenomenon:
 matter of the disc located at large distances
from the star penetrates through the magnetic field of the star
in thin dense filaments. This is a sign of
interchange (or magnetic Rayleigh-Taylor) instability.
Fig. \ref{3dmri-t2-t30} (right panel) also
shows these filaments. The filaments are tall (in the vertical
direction) and narrow (in the radial direction).
   This   shape is favorable for
penetration between the field lines.
     This situation is analogous to penetration of  disc matter
through a vertical magnetic field threading the disc.
      This was  investigated earlier with analytic theory and
 MHD simulations
(\citealt{kais92,  spru95,  lubo95}).
     Here, we observe this
phenomenon in simulations~\footnote{
 Such an instability has been also observed in global 3D MHD simulations by \citet{igum08}, where the magnetic
 flux was accumulated in the inner parts of the disc around the black hole (see also \citealt{nara03}).}.
The interchange instability can be an important mechanism of matter penetration through the magnetic field
in accretion discs in different situations.
     It can give an enhanced transport of magnetic flux.

 Matter of the disc which reaches the disc-magnetosphere boundary, can further penetrate
through the \textit{magnetically-dominated} magnetosphere due to same instability  (e.g., \citealt{aron80, spru90, li04}) or due to the Kelvin-Helmholtz
instability (e.g., \citealt{love10}).
Our earlier global simulations  of $\alpha-$type discs have shown that in magnetized stars  matter flows towards the stellar surface either through two ordered funnel streams, above the
magnetosphere, where two hot spots form, and the light curve is almost sinusoidal. Or, it can penetrate the magnetosphere
due to interchange instability in the
equatorial plane in narrow and tall `tongues'.   In this case, a
number of temporary hot spots form in random places, and the light-curve is chaotic (\citealt{roma08, kulk08}).
This has an important application for understanding of
periodic and quasi-periodic variability in differen stars with small magnetospheres (e.g., \citealt{kulk09, bach10}).

There are several factors which determine the boundary between the stable and unstable regimes of accretion.
The  unstable regime is favorable, if the magnetic axis has a small tilt
about the rotational axis,  if a star rotates slowly (hence, the gravitational plus centrifugal potential is negative),
and also if sufficient amount of matter is accumulated at the disc-magnetosphere boundary.     More, precisely,
the theoretical condition for instability which agrees well with results of simulations
 requires that the gradient of the surface density $\Sigma$ per unit of the vertical magnetic
field $B_z$ should be large and positive, $d(\Sigma/B_z)/dr > 0 $ (e.g., \citealt{spru95,kulk08}).
This term is required to overcome the
stabilizing effect of the velocity shear
(see, e.g.,  eq. 59 of \citealt{spru95}).

In the  present simulations, we see that matter
is lifted above the magnetosphere and accretes in
funnel streams (see Fig. \ref{d10-t2-slice-3}, left panels).
However, this is the case, which is favorable for the
accretion though instabilities:  the star rotates slowly,
 and the tilt $\Theta=2^\circ$ is very small.
        It is possible however that a high disc temperature (see Sec. \ref{sec:init})
 drives matter to the funnel flow and hence opposes to the onset of the instability.
However, this is not the case, because
in the $\alpha-$disc models (e.g., \citealt{roma08}), where instability is observed,
we used the \textit{same parameters} for the density and
temperature distribution in the disc and corona.
     We suggest the instability is not seen in the
current simulations, because the runs are not long enough.
   There is not enough time to build up
 the radial gradient of $(\Sigma/B_z)$ needed
to trigger the instability.
Note also that conditions for the onset of the interchange instability  may be
different in case of a turbulent disc.

\begin{figure*}
\centering
\includegraphics[width=18.0cm]{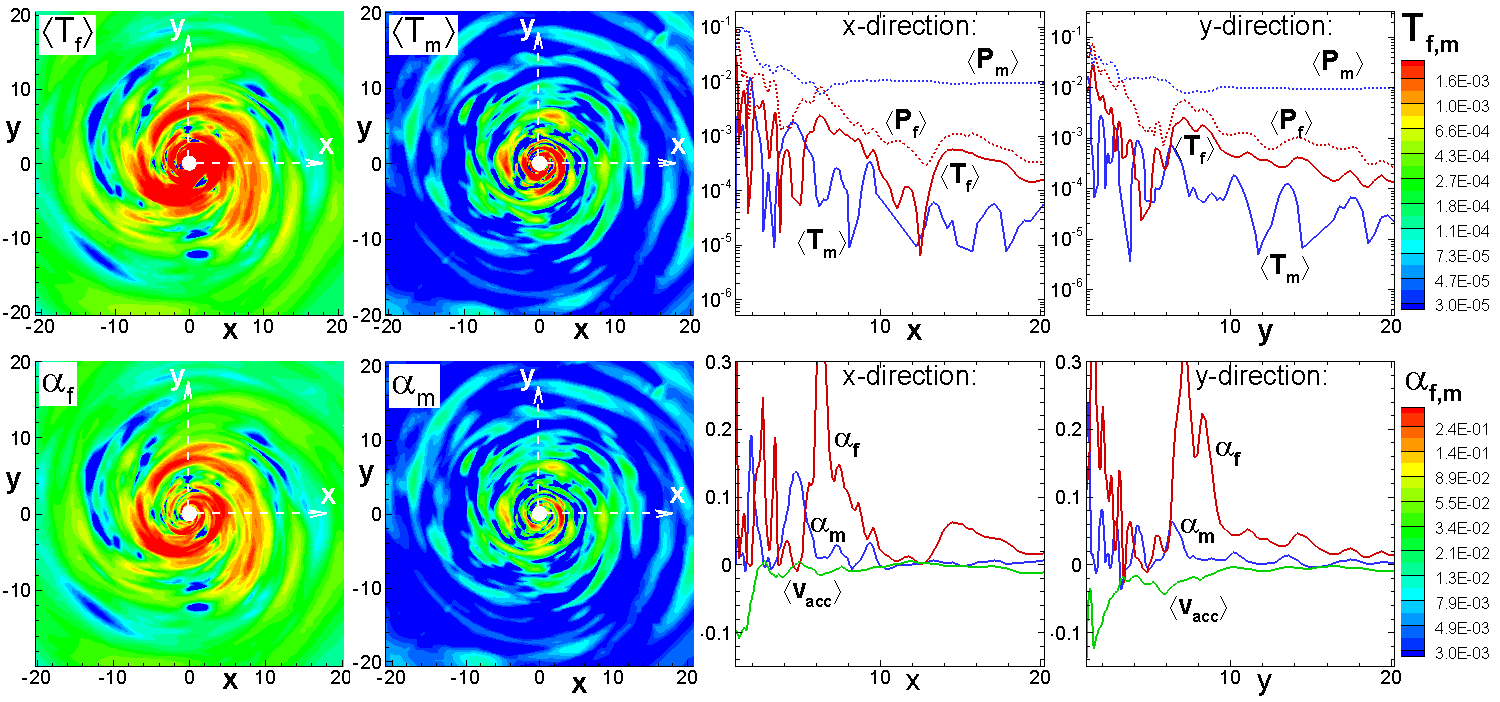}
\caption{The integrated stresses, pressure and $\alpha-$parameters
associated with matter ($\langle{T_m}\rangle$,
$\langle{P_m}\rangle$, $\alpha_m$) and the magnetic field
($\langle{T_f}\rangle$, $\langle{P_f}\rangle$, $\alpha_f$) in the
model with $B'_\star=0.001$ and $\Theta=5^\circ$. \textit{Left two
panels}: the equatorial distribution. \textit{Right two panels:}
distribution along the $x$ and $y$
axes.}\label{d001-t5-stress-8-vac}
\end{figure*}

\begin{figure*}
\centering
\includegraphics[width=18.0cm]{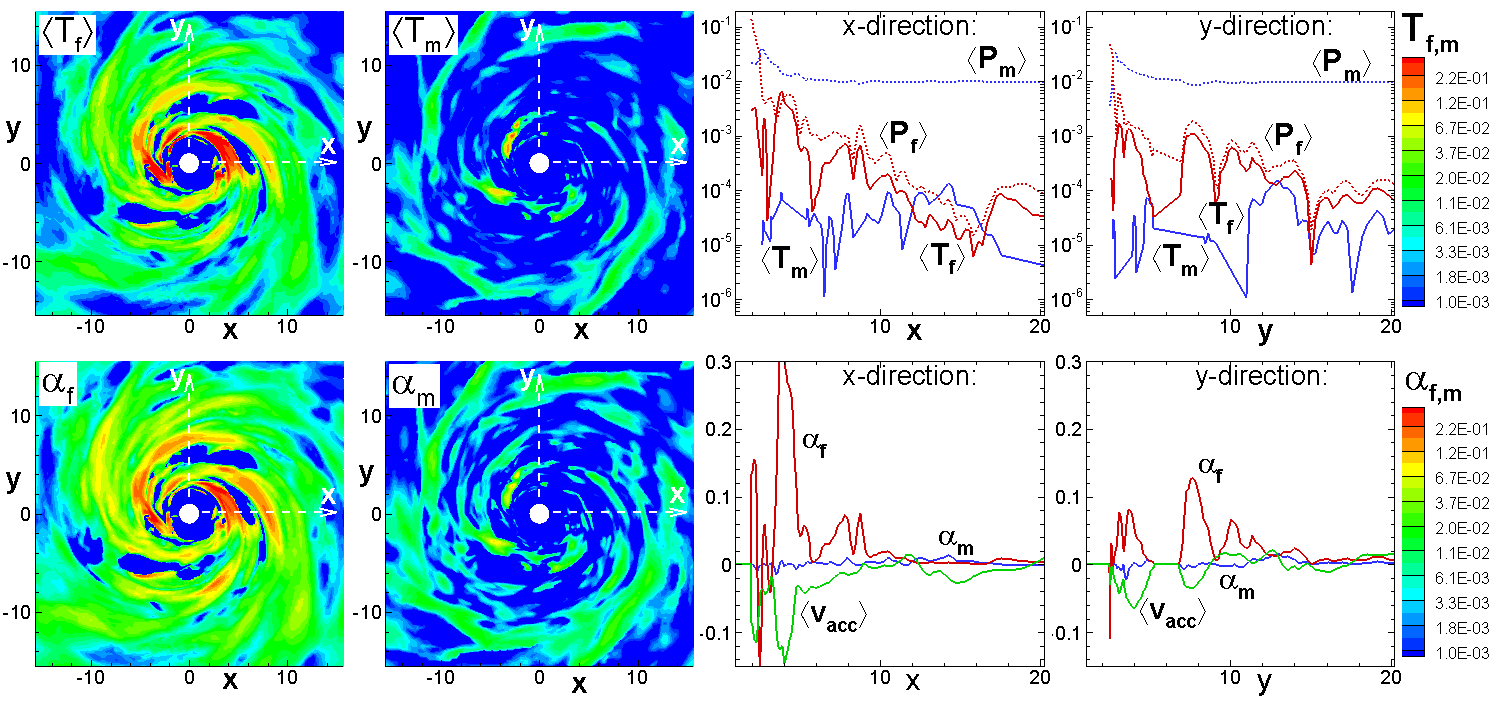}
\caption{Same as in Fig. \ref{d001-t5-stress-8-vac} but for the
case with $B'_\star=10$,
$\Theta=30^\circ$.}\label{d10-t30-stress-8}
\end{figure*}

\section{Analysis of  stresses in the disc}
\label{sec:stresses}

In both cases considered above the turbulent accretion is excited by the magneto-rotational instability.
Here, we analyze stresses in the disc.
    We separate the disc from
the low-density corona using a specific density level,
$\rho_{disc}=0.3$, which is typical for the boundary between the
disc and corona.
     We integrate the matter (subscript `m') and magnetic
(subscript `f') stresses across the disc (in the $z-$direction)
and obtain
$$
\langle{T_m}\rangle = \frac{1}{2h} \int dz \rho v_r v_\phi  -
\langle{\rho v_r}\rangle \langle{v_\phi}\rangle ~,
$$
$$
\langle{T_f}\rangle = - \frac{1}{2h} \int dz \frac{B_r
B_\phi}{4\pi}~.
$$
  Here
$$
\langle{v_\phi}\rangle = \frac{1}{2h} \int dz v_\phi ~,
~~\langle{\rho v_r}\rangle = \frac{1}{2h} \int dz \rho v_r
$$
are averaged azimuthal velocity and matter flux, and $h=h(r)$ is
the half-thickness of the disc.   Note that $r \langle T_m+T_f
\rangle$ is the radial angular momentum flux in the disc.

The matter and magnetic pressures are
$$
\langle{P_m}\rangle = \frac{1}{2h} \int dz P ,
~~~\langle{P_f}\rangle = \frac{1}{2h} \int dz
\frac{{\bf B}^2}{8\pi} .
$$
The standard $\alpha-$parameters associated with matter and
magnetic stresses are:
$$
\alpha_m =
\frac{2}{3}\frac{\langle{T_m}\rangle}{\langle{P_m}\rangle} , ~~~
\alpha_f =
\frac{2}{3}\frac{\langle{T_f}\rangle}{\langle{P_m}\rangle}.
$$

Figure \ref{d001-t5-stress-8-vac} (left-hand four panels) shows
the equatorial distribution of different averaged values in the
disc for the case of the boundary layer accretion, where the magnetic field is relatively weak
($B'_\star=0.001$). One can see that the magnetic stress,
$\langle{T_f}\rangle$, is larger than matter stress,
$\langle{T_m}\rangle$, and hence angular momentum is transported
by the magnetic stress. Note that the matter stress also results
from magnetic turbulence. It is interesting that the magnetic
stress forms a clear one-armed spiral wave. The same wave is seen
as a low-density pattern in Fig. \ref{d001-t5-slice-3-vac}. Slices
for $\alpha-$parameters ($\alpha_f$ and $\alpha_m$) repeat the
pattern for stresses. The four right-hand panels of Fig.
\ref{d001-t5-stress-8-vac} show the  distribution of
stresses along the $x$ and $y$ axes in the equatorial plane.
     One can see that the magnetic stress
is larger on average than the matter stress in both the $x$ and
$y$ directions. Its distribution is very inhomogeneous and it
reflects the distribution of matter and magnetic stresses.
 The $\alpha-$parameters are also strongly
inhomogeneous, and they track the regions of enhanced stresses.
The maximum values are high: $\alpha_f\approx 0.2-0.35$ and
$\alpha_m\approx 0.05-0.18$.

Figure \ref{d10-t30-stress-8} (left-hand four panels) shows the
equatorial distribution of different averaged values in the disc in the case of
the magnetospheric accretion where the magnetic field is relatively strong
and the tilt of the magnetic
moment is high: $B'_\star=10$, $\Theta=30^\circ$. Here, we see that the
equatorial distribution of stresses and $\alpha-$parameters has a
similar pattern.
   The magnetic stress does not repeat the clear
two-armed pattern observed in the density distribution (see Fig.
\ref{d10-t30-slice-3}).

      The matter pressure is much higher than
the magnetic pressure (because the temperature in the disc is
relatively high due to the initial conditions).
However, the magnetic stress is larger than matter stress
at most of radii.

        The bottom left-hand panels show that the magnetic
and matter $\alpha-$parameters ($\alpha_f$ and $\alpha_m$,
respectively) repeat the pattern of the magnetic and matter
stresses. The bottom right-hand panels show that the $\alpha_f$
parameter is largest in spiral waves, ($\alpha_f\approx 0.1-0.3$),
and it is much smaller between them and at larger distances.
            The matter $\alpha-$parameter, $\alpha_m$,
is much smaller than $\alpha_f$.

Note that the simulations in the case of a weak stellar magnetic
field, $B'_\star=0.001$, run longer than in the case of a strong
field. This may explain the formation of clear spiral features in
spite of the small tilt of the magnetic moment, $\Theta=5^\circ$.
In the future, longer simulations runs may show further growth of the
magnetic energy in the disc and more developed spiral structure.

\section{Conclusions and Discussion}

We performed global  three-dimensional ideal MHD simulations of
MRI-driven accretion on to magnetized stars with the dipole
magnetic fields tilted at an angle $\Theta$ relative to the
stellar rotational axis (which coincides with the rotational axis
of the disc). We use the version of our  cubed sphere code
\citep{kold02} where the grid is compressed near the equatorial
plane. Simulations were performed in dimensionless form
and can be applied to different types of stars with relatively
small magnetospheres, such as CTTSs, accreting brown dwarfs, millisecond pulsars,
 and dwarf novae.

Multiple simulation runs were performed for different
parameters of the star and the disc. However, for clarity we show
results for two main regimes of accretion. In one of them, the
magnetic field of the star is dynamically unimportant: it  does
not stop the disc, and matter accretes to the
surface of the star in the BL regime. In another
regime, the magnetic field is dynamically important, it truncates
the disc, and magnetospheric accretion is observed.

Common features observed in both regimes of accretion are the following:


\begin{enumerate}

\item We observed that the MRI-driven turbulence develops in the disc and it
is similar to that observed in cases of accretion on to
non-magnetized objects (black holes) (e.g., \citealt{hawl01}). The magnetic
stress in the disc is larger than the turbulent part of matter stress,
 and hence, the angular momentum is transported outward by the magnetic stress.

\item The disc is strongly inhomogeneous: it consists
of multiple azimuthally-elongated turbulent cells. The azimuthal field is about 10 times larger than
the poloidal field.

\item The spiral structure is observed in the density and stress distributions in the disc.
        It is more
prominent in the case of strongly tilted, dynamically important field
($B'_\star=10$, $\Theta=30^\circ$), and
in case of the dynamically unimportant field
($B'_\star=0.001$, $\Theta=5^\circ$) where simulations are longer.

\smallskip

Below, we show main results for our two regimes of accretion:

\smallskip

\noindent\textit{Boundary Layer regime:}

\smallskip

\item The  magnetic field of the \textit{inner disc} is strongly amplified due to the
shear. The magnetic pressure becomes comparable with
matter pressure, $\beta\approx1-3$, and hence the inner disc is almost in the equipartition.
This result is in accord with earlier simulations by
\citet{armi02} and  \citet{stei02}, who also observed strong amplification of the field in the inner disc.

\item Matter of the disc accretes to the star in individual turbulent cells.
Accretion of the turbulent cell leads to formation of the density `spot' on the star which has a shape of
the equatorial belt. The belt is inhomogeneous and it is wide in the meridional direction.
Spots have lower density and are more narrow during periods of lower accretion rate.


\item The matter flux at the surface of the star varies, and the maxima in the variability curve
correspond to accretion of individual turbulent
cells. The process is quasi-periodic with a period of a few Keplerian rotation
periods  at the radius of the star.

\item The accreting matter brings the poloidal magnetic flux to the star, and a weakly magnetized
star acquires a new, `induced' magnetosphere. The polarity of the induced field depends on the polarity of the disc field,
and hence the field of the star can flip its polarity.

\medskip

\noindent\textit{In the Magnetospheric regime:}

\smallskip

\item The disc is truncated by the magnetosphere of the star at a few stellar radii,
where the magnetic stress in the magnetosphere
balances the total matter stress in the disc.
    Closer to the star matter flows to the star's surface
through the funnel streams.

\item The funnel streams flowing towards the south and north hemispheres are
different, because the accreting turbulent cell flows mainly to the closest magnetic
pole, and less matter flows to another pole. \footnote{In the opposite case of
multiple turbulent cells in the disc (not considered here), the number of cells will
accrete simultaneously and the funnel streams will be more symmetric.}.

\item The hot spots on the star are also different. They reflect the energy distribution in  funnel streams.

\item In the  case of a  small tilt, $\Theta=2^\circ$,
 matter of the disc penetrates through the external magnetosphere of the star due to the interchange instability.
We suggest that this instability can also play an important role in
further penetration of the disc matter through the  closed, magnetically-dominated magnetosphere.
Further work is required to investigate this process.

\end{enumerate}

One of important feature of accretion from turbulent discs (compared with $\alpha-$discs)
is that the turbulence breakes the symmetry between two funnel streams and hot spots in case of magnetospheric
accretion, and it brings non-axisymmetry to the shape of the equatorial belt in case of the boundary layer accretion.
This property is important for understanding the light-curves from magnetized and non-magnetized stars.

The further enhancement of the magnetic field at the disc-star boundary, may possibly lead to the formation of a
magnetic equatorial torus \citep{pacz78} which has been used in the phenomenological models of quasi-periodic oscillation
in dwarf novae cataclysmic variables \citep{warn03,pret06}. On the other hand, the
build up of magnetic field may flow
in to the corona due to the magnetic pressure force  (see, e.g., \citealt{mill00,roma11a}) and it may
be responsible for outflows from young stars (\citealt{prin89}, see also \citealt{shu88}).
Recent axisymmetric simulations show formation of powerful, magnetically-driven outflows from the disc-magnetosphere boundary,
where the strong magnetic field of the star is compressed by the disc at very high accretion rate \citep{lii11}.
These simulations can possibly explain the FU Ori winds as discussed by \citet{koni11}.
We have not seen formation of fast outflows
in  current simulations of accretion in the BL regime. Recently, we were able to obtain outflows from stars rotating in
the `propeller' regime (Ustyugova et al. 2011). However, this is the case
of the dynamically important field.

An interesting phenomenon of magnetospheric accretion through interchange instability observed in modelling of
$\alpha-$discs  (e.g., \citealt{roma08,kulk08}), has not been observed in current simulations.
We suggest that our simulations of magnetospheric accretion were not long enough to observe this instability, and/or
 the instability criterion can be somewhat different for cases of turbulent and $\alpha-$type discs.

\section*{Acknowledgments}

Resources supporting this work were provided by the NASA High-End Computing (HEC)
Program through the NASA Advanced Supercomputing (NAS) Division at Ames Research Center and
the NASA Center for Computational Sciences (NCCS) at Goddard Space Flight Center.
   The research was supported by NASA grants
NNX10AF63G, NNX11AF33G and NSF grant AST-1008636.
AVK and GVU were supported in part by RFBR grants 09-01-00640a and
09-02-00502a.


\end{document}